\documentclass[conference,letter,10pt]{IEEEtran}

\usepackage{amsmath,epsfig,cite,amsfonts,amssymb,psfrag,subfig}

\def\no{\nonumber}

\newtheorem{theorem}{Theorem}

\newtheorem{col}{Corollary}

\newtheorem{rem}{Remark}
\newtheorem{defi}{Definition}

\begin{document}
%don't want date printed
%\date{}
%make title bold and 14 pt font (Latex default is non-bold, 16pt)
\title{Pairwise Secret Key Agreement based on Location-derived Common Randomness }
%for single author (just remove % characters)
%\author{I. M. Author \\
%  My Department \\
%  My Institute \\
%  My City, STATE, zip}
%for two authors (this is what is printed)

\author{\IEEEauthorblockN{Somayeh Salimi, Panos Papadimitratos}
\IEEEauthorblockA{Networked Systems Security Group, School of Electrical Engineering, KTH, Stockholm, Sweden\\
somayen@kth.se,  papadim@.kth.se}
}
%\author{Somayeh Salimi}
%, Mikael Skoglund,{tabular}[t]{c@{\extracolsep{8em}}c}
%I. M. Author  & M. Y. Coauthor \\
% \\ACCESS Linnaeus Center, School of Electrical Engineering, KTH
% \\Stockholm, Sweden \\{ somayen@kth.se,skoglund@ee.kth.se}

%I don't know why I have to reset thispagestyle, but otherwise get page numbers
%\thispagestyle{empty}
%\subsection*{\centering Abstract}
%IEEE allows italicized abstract
%{\em

\maketitle

\begin{abstract}
A source model of key sharing between three users is considered in
which each pair of them wishes to agree on a secret key hidden
from the remaining user. There are rate-limited public channels for communications between the users. We give an inner bound on the secret key capacity region in this framework.
Moreover, we investigate a practical setup in which localization information of the users as the correlated observations are exploited to share pairwise keys between the users. The inner and outer bounds of the key capacity region are analyzed in this setup for the case of i.i.d. Gaussian observations.
\end{abstract}

%
%% How to insert a fig
%\begin{figure}%[b]
%    %\centering
%%\begin{center}
%    %\psfrag{fy}[][][2.2]{ $f(Y_r)$ }
%    %\resizebox{8cm}{!}{\epsfbox{Drawing2.eps}}
%    %\vspace*{-1.7cm}
%    \caption{Block diagram of the two-hop  relay channel.}\label{fig:2hop_relay}
%    %\vspace*{-.4cm}
%%\end{center}
%\end{figure}

\vspace{-.2cm}
%\footnote{This work is partially supported by.}
\section{Introduction}
\noindent Secret key sharing at the physical layer is a promising approach for deriving shared secret keys. Ahlswede and Csiszar \cite{Ahlswede} and Maurer \cite{Maurer} introduced source and channel models of key sharing between two legitimate users in the presence of an eavesdropper using source and channel common randomness along with an unlimited public channel. Various extensions considered a limited public channel \cite{Csiszar}, sharing of one secret key in a network of users \cite{multiterminal}, and more than one secret key in different scenarios \cite{threeter}--\cite{salimi-pairwise}.

\emph{Pairwise key sharing} first introduced in \cite{salimi-pairwise}, is a specific problem in this area, requiring that each pair of users shares a secret key concealed from the remaining user(s). In a basic setup including three users with access to correlated source observations and communication over an unlimited public channel, inner and outer bounds on the secret key capacity region were derived. In this paper, we extend the pairwise key sharing framework in \cite{salimi-pairwise} to the rate-limited public channel for communications. The public channel is full duplex and each of the users can simultaneously send/receive information over/from the public channel. Based on the correlated observations, users communicate over the rate-limited public channel. Then, each user generates the respective keys as functions of its source observations and the information received over the rate-limited public channel. We derive an inner bound on the key capacity region in this framework; the explicit outer bound given in \cite{salimi-pairwise} holds here for the rate-limited public channel case.

We consider location-derived common randomness here because it is a promising, towards practical applications, approach. This is so because a multitude of emerging wireless systems are location-aware and devices can and need to perform distance measurements over RF communication, notably for security reasons, for example \cite{SND},\cite{TMC}.

Location-derived common randomness was considered in \cite{PFU} in a different setup, with a key established between a mobile node and a wireless infrastructure. In a setup closer to the one considered here, \cite{mobility-conf} considered two users that move according to a discrete time stochastic mobility model and measure their respective distance, after exchanging messages, in the presence of an eavesdropper. In this paper, leveraging the latter approach, we generalize location-derived key sharing to the \lq\lq pairwise secret key\rq\rq setting, notably with three users. We present inner bounds of the pairwise key capacity region for both unlimited and limited public channels. Furthermore, the explicit outer bound in \cite{salimi-pairwise} is analyzed in this i.i.d. Gaussian setup. Some numerical results are given for the Gaussian setup as well.

The proposed scheme can be extended to the case of more than three users as the future work in which collusion of curious users needs to be investigated. Here we consider simply users curious about the keys their peers derive. But they do not otherwise deviate from the specification and disrupt the protocol.

%If the users don't collude to eavesdrop, the eavesdropper of each pair's key is the strongest among the remaining users, i.e., the user who has the best estimation of the distance between this pair. If the eavesdroppers collude, they combine their observation to make a better estimation of the distance between an intended pair.

The rest of the paper is organized as follows: in Section II, the preliminaries of the key sharing setup are given. An inner bound of the pairwise key capacity region with rate-limited public channel is given in Sections III. Deriving pairwise keys from localization information along with the respective inner and outer bounds are presented in Section IV. Numerical results and concluding remarks are given in Sections V and VI, respectively. Proofs of the results are presented in Appendices.

\begin{figure}%[!t]
\centering
%\vspace{-2.5cm}
\includegraphics[width=6cm]{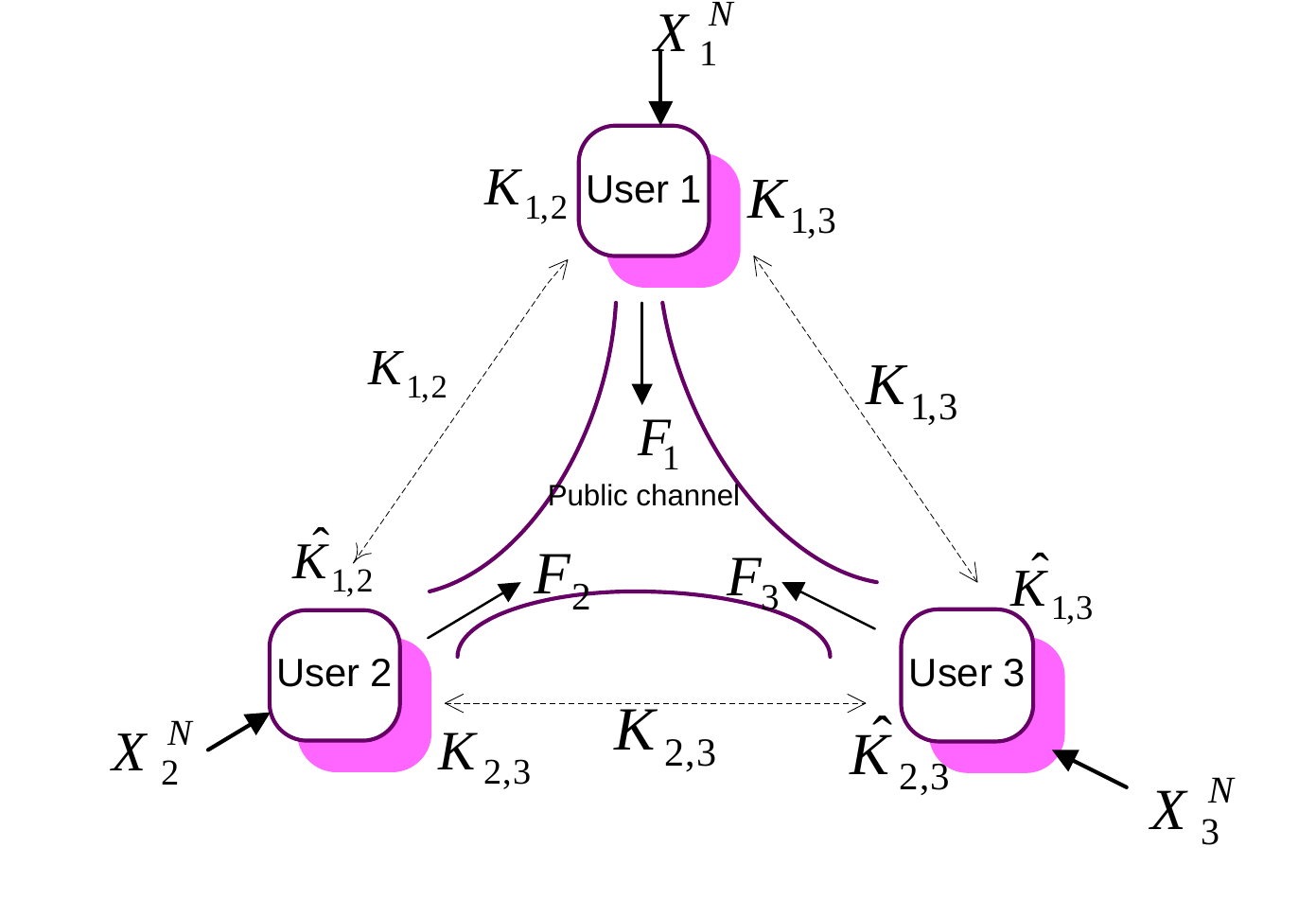}
%\vspace{-2.5cm}
\footnotesize{
\caption{Pairwise secret key sharing in the source model}
}
\vspace{-.5cm}
\label{fig_sim}
\end{figure}

\normalsize{
\section{Preliminaries}
\vspace{-.1cm}
 \noindent Users 1, 2 and 3, respectively, have access to $n$ i.i.d. observations $X_{1}, X_{2}$ and $X_{3}$ according to Fig. 1. The observations are correlated according to distribution $P_{X_{1}X_{2}X_{3}}$. The random variable $X_{i}$ takes values from the finite set ${\mathcal X}_{i}$ for $i=1,2,3$. Furthermore, there exists a noiseless public channel of limited capacity for communication between the three users where user $i$ is subject to rate constraint $R_{i}$ for its transmission. Each pair of the three users intends to share a secret key concealed from the remaining user. $K_{i,j}$ denotes the shared key between users $i$ and $j$, hidden from user $m$, for $i,j,m\in\{1,2,3\}$, $i<j$, $m\neq i,j$. We represent the formal definition of the described secret key sharing setup.
 }

%We assume each user once sends information over the limited-rate public channel.
User $i$ sends stochastic function $F_{i}=f_{i}(X_{i}^{n})$ over the rate-limited public channel for $i=1,2,3$ subject to
\begin{equation}
\frac{1}{n}H(F_{i})\leq R_{i}
\end{equation}
Upon receiving the information over the public channel, key generation is performed at the users. Key generation function $g_{i}$ is used by user $i$ for $i=1,2,3$ as:
\begin{align}
    g_{1}&:\mathcal{F}_2\times\mathcal{F}_3\times\mathcal{X}_1^n\rightarrow\mathcal{K}_{1,2}\times\mathcal{K}_{1,3} \\
    g_{2}&:\mathcal{F}_1\times\mathcal{F}_3\times\mathcal{X}_2^n\rightarrow\mathcal{K}_{1,2}\times\mathcal{K}_{2,3} \\
    g_{3}&:\mathcal{F}_1\times\mathcal{F}_2\times\mathcal{X}_3^n\rightarrow\mathcal{K}_{1,3}\times\mathcal{K}_{2,3}.
  \end{align}

Thus, user 1 calculates $K_{1,2}$ and $K_{1,3}$ to share with users 2 and 3, respectively. Similarly, user 2 calculates $\hat{K}_{1,2}$ and $K_{2,3}$ to share with users 1 and 3 and user 3 calculates $\hat{K}_{1,3}$ and $\hat{K}_{2,3}$ to share with users 1 and 2.

%\textbf{\textit{Definition 1}}:
\begin{defi}
In the pairwise secret key sharing over public channels of limited rates $(R_{1},R_{2},R_{3})$ at the respective users 1, 2, 3, the rate triple $(R_{12},R_{13},R_{23})$ is an achievable key rate pair if for every $\varepsilon>0$ and sufficiently large $n$, we have:
\begin{align}
&\forall i<j\in\{1,2,3\}\quad \frac{1}{n}H(K_{i,j})\geq R_{ij}-\epsilon \label{eq1}\\
&\forall i<j\in\{1,2,3\}\quad\Pr\{K_{i,j}\neq\hat{K}_{i,j}\}<\varepsilon \label{eq2}\\
&\forall i\!<\!j,m\!\in\!\{1,2,3\},\!m\notin\!\{i,j\}\ \ I(K_{i,j};F_{i},F_{j},X_m^n)<\varepsilon  \label{eq3}\\
&\forall i\!\in\!\{1,2,3\}\ \ \ \ \frac{1}{n}H(F_{i})\leq R_{i}\label{eq4}.
 \end{align}
%\noindent
%\[\begin{array}{l} {\begin{array}{c} {\forall i<j\in\{1,2,3\}{\tfrac{1}{n}} H(K_{i,j} )>R_{ij} -\varepsilon} \\ {{\tfrac{1}{n}} H(K_{1,3} )>R_{13} -\varepsilon}
%\\ {{\tfrac{1}{n}} H(K_{2,3} )>R_{23} -\varepsilon } \end{array}{\rm\; \; \; \; \; \; \; \; \; \; \; \; \; \; \; \; \; \; \; \; \; \; \; \; \; \; \; \; \; \; \; \; \; \; \; \; \; \; }\left(\begin{array}{c} {}
%\\ {4} \\ {} \end{array}\right)}
%\\ {\begin{array}{c} {\Pr \{ K_{1,2} \ne \hat{K}_{1,2} \} <\varepsilon} \\ {\Pr \{ K_{1,3} \ne \hat{K}_{1,3} \} <\varepsilon} \\ {\Pr \{ K_{2,3} \ne \hat{K}_{2,3} \} <\varepsilon} \end{array}{\rm\;\; \;  \; \; \; \; \; \; \; \; \; \; \; \; \; \; \; \; \; \; \; \; \; \; \; \; \; \; \; \; \; \; \; \; \; \; \ }\left(\begin{array}{c} {} \\ {5} \\ {} \end{array}\right)} \\ {{\tfrac{1}{n}} I(K_{1,2} ;X_{3}^{n} ,F_{1} ,F_{2} )<\varepsilon {\rm \; \; \; \; \; \; \; \; \; \; \; \; \; \; \; \; \; \; \; \; \; \; \; \; \; \; \; \; \; \;\; \; \; \; \; \; \; \; \;(6)}} \\ {{\tfrac{1}{n}} I(K_{1,3} ;X_{2}^{n} ,F_{1} ,F_{3} )<\varepsilon {\rm\; \; \;\; \; \; \; \; \; \; \; \; \; \; \; \; \; \; \; \; \; \; \; \; \; \; \; \; \; \; \; \; \; \; \; \; \; \; \;(7)}} \\ {{\tfrac{1}{n}} I(K_{2,3} ;X_{1}^{n} ,F_{2} ,F_{3} )<\varepsilon. {\rm \; \; \; \; \; \; \; \; \; \; \; \; \; \; \; \; \; \; \; \; \; \; \; \; \; \; \; \; \; \; \; \; \; \; \; \; \; \; (8)}} \end{array}\]
\end{defi}

\noindent\normalsize{Equation \eqref{eq1} means that rate $R_{ij}$ is the rate of the secret key between users $i$ and $j$.
Equation \eqref{eq2} means that each user can correctly estimate the respective
keys. Equation \eqref{eq3} means that each user effectively
has no information about the remaining users' secret key. Equation \eqref{eq4} denotes that the key sharing is subject to the constraint of the public channel.

\begin{defi}
The region containing the entire
achievable secret key rate triples $(R_{12},R_{13},R_{23})$ is
the secret key capacity region.
\end{defi}
\vspace{-.4cm}

\section{ Main Result }
\vspace{-.2cm}
\noindent In the following, an inner bound on the pairwise key capacity region of the source model with rate-limited public channel is given. First, we define:
%\begin{equation} \label{eq:test1}
\[\begin{array}{l} {{\bf r}_{{\bf 12}} =[I(S_{12} ;X_{2} \left|S_{23} S_{32} )\right. -I(S_{12} ;X_{3} ,S_{13} \left|S_{23} ,S_{32} )\right. ]^{+} ,}
\\{{\bf r}_{{\bf 21}} =[I(S_{21} ;X_{1} \left|S_{13} S_{31} )\right. -I(S_{21} ;X_{3} ,S_{23} \left|S_{13} ,S_{31} \right. )]^{+} ,}
\\ {{\bf r}_{{\bf 13}} =[I(S_{13} ;X_{3} \left|S_{23} S_{32} )\right. -I(S_{13} ;X_{2} ,S_{12} \left|S_{23} ,S_{32} )\right. ]^{+} ,}
\\{{\bf r}_{{\bf 31}} =[I(S_{31} ;X_{1} \left|S_{12} S_{21} )\right. -I(S_{31} ;X_{2} ,S_{32} \left|S_{12} ,S_{21} \right. )]^{+} ,}
\\ {{\bf r}_{{\bf 23}} =[I(S_{23} ;X_{3} \left|S_{13} S_{31} )\right. -I(S_{23} ;X_{1} ,S_{21} \left|S_{13} ,S_{31} )\right. ]^{+} ,}
\\ {{\bf r}_{{\bf 32}} =[I(S_{32} ;X_{2} \left|S_{12} S_{21} )\right. -I(S_{32} ;X_{1} ,S_{31} \left|S_{12} ,S_{21} \right. )]^{+} },
\\ {{\bf I}_{{\bf 12}} =I(S_{12};S_{21} \left|X_{3} ,S_{13} ,S_{23} )\right. },\\{{\bf I}_{{\bf 13}} =I(S_{13} ;S_{31} \left|X_{2} ,S_{12} ,S_{32} )\right.},
\\ {{\bf I}_{{\bf 23}} =I(S_{23} ;S_{32} \left|X_{1} ,S_{21} ,S_{31} )\right. },{{\bf I}_{{\bf 1}} =I(S_{21} ;S_{31} \left|X_{1} )\right.},
 \\{{\bf I}_{{\bf 2}} =I(S_{12} ;S_{32} \left|X_{2})\right. },{{\bf I}_{{\bf 3}} =I(S_{13} ;S_{23} \left|X_{3} )\right.}.
\end{array}\]
%\end{equation}

%in Equation \eqref{eq:test1}.
%\begin{eqnarray}
%{\bf r}_{{\bf 12}} =[I(S_{12} ;X_{2} \left|S_{23} S_{32} )\right. -I(S_{12} ;X_{3} ,S_{13} \left|S_{23} ,S_{32} )\right. ]^{+} ,
%\\{\bf r}_{{\bf 21}} =[I(S_{21} ;X_{1} \left|S_{13} S_{31} )\right. -I(S_{21} ;X_{3} ,S_{23} \left|S_{13} ,S_{31} \right. )]^{+} ,
%\\ {\bf r}_{{\bf 13}} =[I(S_{13} ;X_{3} \left|S_{23} S_{32} )\right. -I(S_{13} ;X_{2} ,S_{12} \left|S_{23} ,S_{32} )\right. ]^{+} ,
%\end{eqnarray}
\vspace{-.2cm}
%\textbf{\textit{Theorem 1 (inner bound)}}:
\begin{theorem}\label{th1}
In the described setup, all rates in the closure of the convex hull of the set of all key rate triples $(R_{12},R_{13},R_{23})$ that satisfy the following region, are achievable:
%\vspace{.3cm}
\begin{align}
&{R_{12} >0,R_{13} >0,R_{23} >0,}\no
\\& {R_{12} \le {\bf r}_{{\bf 12}} {\bf +r}_{{\bf 21}} -{\bf I}_{{\bf 12}} ,}\no
\\&{R_{13} \le {\bf r}_{{\bf 13}} {\bf +r}_{{\bf
31}} -{\bf I}_{{\bf 13}},}\no
\\&{R_{23} \le {\bf r}_{{\bf 23}} {\bf +r}_{{\bf 32}} -{\bf I}_{{\bf 23}},}\no
\\&{R_{12} +R_{13}\le {\bf r}_{{\bf 12}} {\bf +r}_{{\bf 21}} {\bf +r}_{{\bf 13}}
{\bf +r}_{{\bf 31}}-{\bf I}_{{\bf 12}}-{\bf I}_{{\bf 13}}
-{\bf I}_{{\bf 1}},}\no
 \\&{R_{12} +R_{23} \le{\bf r}_{{\bf 12}} {\bf +r}_{{\bf 21}} {\bf +r}_{{\bf 23}} {\bf
+r}_{{\bf 32}} -{\bf I}_{{\bf 12}}-{\bf I}_{{\bf 23}}-
{\bf I}_{{\bf 2}},}\no
\\&{R_{13} +R_{23} \le{\bf r}_{{\bf 13}} {\bf +r}_{{\bf 31}} {\bf +r}_{{\bf 23}} {\bf
+r}_{{\bf 32}} -{\bf I}_{{\bf 13}}-{\bf I}_{{\bf 23}}
-{\bf I}_{{\bf 3}},}\no
\\&{R_{12} +R_{13} +R_{23} \le {\bf r}_{{\bf 12}}
{\bf +r}_{{\bf 21}} {\bf +r}_{{\bf 13}} {\bf +r}_{{\bf 31}} {\bf
+r}_{{\bf 23}} {\bf +r}_{{\bf 32}} -}\no
\\&{\ \ \ \ \ \ \ \ \ \ {\bf I}_{{\bf 12}}-{\bf I}_{{\bf 13}}-{\bf I}_{{\bf 23}}
-{\bf I}_{{\bf 1}}-{\bf I}_{{\bf 2}}-{\bf I}_{{\bf 3}}}\label{rateregion}
\end{align}
%\vspace{.3cm}

\noindent for random variables taking values in sufficiently large finite sets and according to the distribution:
\[\begin{array}{l} {{\rm \; \; \; \; \; \;
\; \; \; }p(s_{12} ,s_{13},s_{21} ,s_{23} ,s_{31} ,s_{32} ,x_{1} ,x_{2} ,x_{3} )=p(x_{1}
,x_{2} ,x_{3} ). } \\ {p(s_{12}|x_{1})p(s_{13}|x_{1})p(s_{21}|x_{2} )p(s_{23}|x_{2} )p(s_{31}|x_{3} )p(s_{32}|x_{3} )} \end{array}\]
and subject to the constraints:
%\end{align}
\small{
\begin{align}
&{I\!(\!S_{12};\!X_{\!1}\!|X_{2},\!S_{32}\!)\!+\!\!I\!(\!S_{13};\!X_{\!1}\!|X_{3},\!S_{23}\!)\!\leq\! R_{1},}\label{cons1}
\\&{I\!(\!S_{21};\!X_{\!2}\!|\!X_{\!1},\!S_{31}\!)\!+\!I\!(\!S_{23};\!X_{\!2}\!|\!X_{3},\!S_{13}\!)\!\leq\! R_{2},}\label{cons2}
\\&{I\!(\!S_{31}\!;X_{\!3}\!|\!X_{\!1},\!S_{21}\!)\!+\!I\!(\!S_{32};\!X_{\!3}\!|\!X_{\!2},\!S_{12}\!)\leq\! R_{3},}\label{cons3}
\hspace{-.1cm}\\&{I(\!S_{12};\!X_{\!1}\!|X_{\!2},\!S_{32}\!)\!+\!I(\!S_{21};\!X_{\!2}\!|X_{\!1},S_{31}\!)\!+\!I(S_{13},\!S_{23};\!X_{\!1},\!X_{\!2}\!|X_{\!3}\!)}\no
\\&{\leq R_{1}+R_{2},}\label{cons12}
\\&{I\!(\!S_{13};\!X_{\!1}\!|X_{\!3},\!S_{23}\!)\!+\!I\!(\!S_{31};\!X_{\!3}\!|X_{\!1},\!S_{21}\!)\!+\!I\!(\!S_{12},\!S_{32};\!X_{\!1},\!X_{\!3}\!|X_{\!2}\!)}\no
\\&{\leq R_{1}+R_{3},}\label{cons13}
\\&{I\!(\!S_{23};\!X_{\!2}\!|X_{\!3},\!S_{13}\!)\!+\!I\!(\!S_{32};\!X_{\!3}\!|X_{\!2},\!S_{12}\!)\!+\!I\!(\!S_{21},\!S_{31};\!X_{\!2},\!X_{\!3}\!|X_{\!1}\!)}\no
\\&{\leq R_{2}+R_{3}.}\label{cons23}
\\&{\!I\!(\!S_{21},\!S_{31};\!X_{\!2},\!X_{\!3}\!|X_{\!1}\!)+\!I\!(\!S_{12},\!S_{32};\!X_{\!1},\!X_{\!3}\!|X_{\!2}\!)+\!I(S_{13},\!S_{23};\!X_{\!1},\!X_{\!2}\!|X_{\!3})}\no
\\&{\leq R_{1}+R_{2}+R_{3}.}\label{cons123}
\end{align}}
\end{theorem}

\normalsize{
\begin{IEEEproof}
The proof of Theorem \ref{th1} is given in Appendix A.
\end{IEEEproof}}

The rate region in Theorem \ref{th1} is achieved by double random binning as well as Wyner-Ziv coding \cite{Wyner-Ziv} and rate splitting. In the achievability scheme, the rate of the key between users $i$ and $j$ consists of two parts. A part is rate of the key generated by user $i$ to share with user $j$ (${\bf r}_{ij}$) and the other part is the rate of the key generated by user $j$ to share with user $i$ (${\bf r}_{ji}$). The auxiliary random variable $S_{ij}$ stands for the former key while $S_{ji}$ is associated with the latter key. The total rate of the key between users $i$ and $j$ is the sum of ${\bf r}_{ij}$ and ${\bf r}_{ji}$ in which term $\textbf{I}_{ij}$ is subtracted to avoid revealing any information about one of the key to the remaining user (as the eavesdropper) in the case that the other key is disclosed. The limitation of the public channel at the users is reflected in \eqref{cons1}-\eqref{cons123}.

%\begin {rem}
%We can assume that each of the users sends part of the required information by transmitting the bin index of an auxiliary random variable directly over the public channel like \cite{Ahlswede}, however to avoid the complexity of the equations in the secret key rate region, we relinquish it.\textbf{\textit{ }}
%\end {rem}

\begin {rem}
The region in Theorem \ref{th1} reduces to key rate regions in \cite{salimi} by considering subset of keys and assuming unlimited public channel. It also reduces to the key rate region in \cite{salimi-pairwise} by removing public channel limitations.
\end {rem}

%\begin {rem}
% If we assume that two pairs of the users including users 1 and 2, and users 1 and 3 intend to share secret keys and users 2 and 3 are allowed to transmit, our obtained region reduces to the secret key rate region of the forward situation in \cite{salimi} by substituting $S_{12}=S_{13}=S_{23}=S_{32}=\phi$ in Theorem 1.\textbf{\textit{}}
%\end {rem}
%
%\begin {rem}
%If we assume that two pairs of the users including users 1 and 2, and users 1 and 3 intend to share secret keys and user 1 is allowed to transmit, our obtained region reduces to the secret key rate region of the backward situation in \cite{salimi} by substituting $S_{21}=S_{31}=S_{23}=S_{32}=\phi$ in Theorem 1.\textbf{\textit{}}
%\end {rem}
We do not present a new outer bound on the key capacity region. The explicit outer bound in \cite{salimi-pairwise} with unlimited public channel holds in this new setup.

%\begin{theorem}
%In the described setup of pairwise secret key sharing, the following bound is an explicit upper bound:
%\[\begin{array}{l} {R_{12} \le I(X_{1} ;X_{2} \left|X_{3} )\right. ,} \\ {R_{13} \le I(X_{1} ;X_{3} \left|X_{2} )\right. ,} \\ {R_{23} \le I(X_{2} ;X_{3} \left|X_{1} )\right. .} \end{array}\]
%The proof of Theorem 2 is relinquished since it can be considered as a extension of the explicit outer bound in \cite{Ahlswede} and \cite{Maurer} for three secret keys.
%
%\end{theorem}
\section{A Real-World Example of the Pairwise Key Sharing}
In this section, we consider pairwise key sharing between three users who move in two-dimensional space according to a discrete time stochastic mobility model. The idea of using localization information to share a secret key between two users in the presence of an eavesdropper was first introduced in \cite{mobility-conf}. Here, we extend this idea to the pairwise key sharing between three users. The users are mobile in continuous space according to a discrete time stochastic mobility model, independent of each  other. Each pair of the three mobile users exploit the distance between themselves as a source of common randomness to share a key while the remaining user tries to make an estimate of that distance as precise as possible. We borrow some notations from \cite{mobility-conf}. We assume the considered time is divided into $n$ discrete time slots where time slot $l$ includes the time interval $[lT,(l+1)T]$. The users' locations are assumed constant during a time slot. As shown in Fig. 2, at time slot $l$, the distance between users $i$ and $j$ is $d_{ij}[l]=|x_{i}[l]-x_{j}[l]|$ in which $x_{i}[l] \in \mathbb{R}^{2}$ is the random variable which denotes user $i$ location at time slot $l$. In the same figure, $\phi_{i}[l]$ shows the angle of the triangle at user $i$ at time slot $l$. Each pair first exchanges beacon signals (e.g., using propagation delay) to make correlated observations and then, they communicate over the (limited) public channel to share a key hidden from the remaining user. This is performed in two phases as follow.
\begin{figure}%[!t]
\centering
\vspace{-.5cm}
\includegraphics[width=4cm]{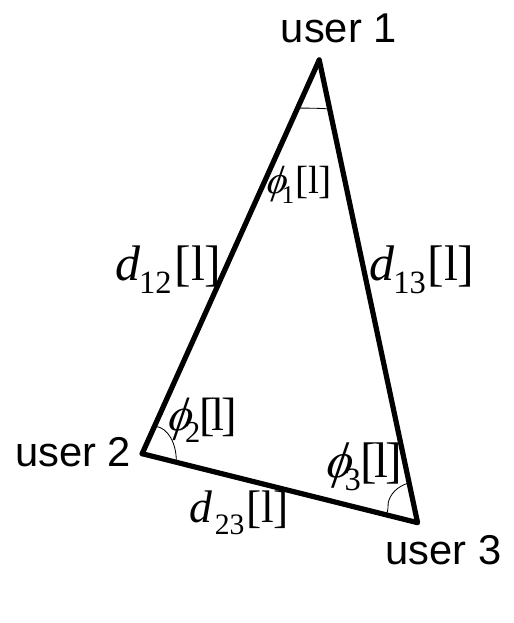}
%\vspace{-2.5cm}
\vspace{-.5cm}
\footnotesize{
\caption{Using location information for Pairwise secret key sharing}
}
\vspace{-.5cm}
\label{fig_sim}
\end{figure}

\textbf{Localization phase:} User $i$ broadcasts some beacons (as a short signal
bearing localization information on the initiating node) at the beginning of time slot $l$ and users $j$ and $m$ obtain noisy observations of $d_{ji}[l]$ and $d_{mi}[l]$, respectively, for $i \in\{1,2,3\}, j\neq m \in \{1,2,3\}-i$. We assume the users are equipped to directional antenna and hence, user $i$ obtain $\hat{\phi}_{i}[l]$ as the noisy version of the angle between the remaining two users. The same as in \cite{mobility-conf}, we assume the sent information by the users is corrupted by Gaussian noises. We have:
\begin{align}
&\tilde{d}_{ij}[l]=d_{ij}[l]+N_{ij}[l] \label{dij}\\
&\tilde{\phi}_{i}[l]=\phi_{i}[l]+N_{i}[l]\label{phii}
 \end{align}
\noindent where $N_{ij}[l]$ and $N_{i}[l]$ are zero-mean Gaussian noises with variances $\sigma_{ij}^{2}$ and $\sigma_{i}^{2}$, respectively. All the noises are independent of each other. In the rest of the paper, we consider the case of i.i.d. locations and additive noises. Thus, we drop index $l$ in equations \eqref{dij}-\eqref{phii}. If the number of broadcast beacons by each user is $J\geq 1$, then $\sigma_{ij}^{2}$ and $\sigma_{i}^{2}$ are divided by $J$ \cite{mobility-conf}. We assume that users are perfectly clock synchronized (it is shown in \cite{mobility-conf} that clock mismatch does not affect the theoretical bounds of secret key rates).

\textbf{Key generation by public channel communications:} At the beginning of this phase, user $i$ has access to its observations
\begin{equation}
\textbf{o}_{i}\!=\!\{\tilde{\textbf{d}}_{ij}\!=\!\{\tilde{d}_{ij}[l]\}_{l=1}^{n}, \tilde{\textbf{d}}_{im}\!=\!\{\tilde{d}_{im}[l]\}_{l=1}^{n},\tilde{\boldsymbol{\phi}}_i\!=\!\{\tilde{\phi}_i[l]\}_{l=1}^{n}\}\label{useriobservation}
\end{equation}
The users communicate over a (rate-limited) public channel to share secret keys in the pairwise manner. Users $i$ and $j$ exploit the reciprocity of the distance between themselves to share a key based on their noisy observations $\tilde{\textbf{d}}_{ij}$ and $\tilde{\textbf{d}}_{ji}$, respectively:
\begin{align}
&\tilde{d}_{ij}=d_{ij}+N_{ij}\label{d12}\\
&\tilde{d}_{ji}=d_{ji}+N_{ji}\label{d21},
 \end{align}
where $d_{ij}=d_{ji}$ is the real distance and $N_{ij}\sim \mathcal{N}(0,\sigma_{ij}^{2}/J)$, $N_{ji}\sim \mathcal{N}(0,\sigma_{ji}^{2}/J)$ assuming each user broadcasted $J$ beacons at the localization phase. On the other hand, the remaining user $m$ tries to estimate $d_{ij}$ to obtain information about the key between users $i$ and $j$ as much as possible with access to $(\tilde{d}_{mi},\tilde{d}_{mj},\tilde{\phi}_{m})$.

Due to simplicity, we assume $\sigma_{ij}=\sigma_{ji}$ between each pair $i$ and $j$. In continue, we consider unlimited and rate-limited public channels separately.
\subsection{unlimited public channel}\label{unlimitedpc}
Since the observation between pair $i$ and $j$ is symmetric (because of $\sigma_{ij}=\sigma_{ji}$) and the public channels at both sides are unlimited, we choose one-way communication between each pair. Without loss of generality, it is assumed that user 1 communicates to user 2, user 2 communicates to user 3 and user 3 communicates to user 1.
According to the directions of communications between users, we choose $S_{12}=\tilde{d}_{12},S_{23}=\tilde{d}_{23},S_{31}=\tilde{d}_{31},S_{21}=S_{32}=S_{13}=\phi$ in Theorem 1. Then the rate region in Theorem 1 is reduced to:
\begin{align}
&{R_{12} >0,R_{13} >0,R_{23} >0,}
\\&{R_{12}\leq I\!(\!\tilde{d}_{12};\!\tilde{d}_{\!21}\!)\!-\!\!I\!(\!\tilde{d}_{12};\tilde{d}_{\!31},\tilde{d}_{\!32},\tilde{\phi}_{3}\!)}\label{apprate12}
\\&{R_{13}\leq I\!(\!\tilde{d}_{31};\!\tilde{d}_{\!13}\!)\!-\!\!I\!(\!\tilde{d}_{31};\tilde{d}_{\!21},\tilde{d}_{\!23},\tilde{\phi}_{2}\!)}\label{apprate13}
\\&{R_{23}\leq I\!(\!\tilde{d}_{23};\!\tilde{d}_{\!32}\!)\!-\!\!I\!(\!\tilde{d}_{23};\tilde{d}_{\!12},\tilde{d}_{\!13},\tilde{\phi}_{1}\!)}\label{apprate23}
\end{align}
Each potential eavesdropper combines its available observations to estimate the distance between the other two users to enlarge the subtracted mutual information terms in \eqref{apprate12}-\eqref{apprate23}. Thus, user $m$ as a potential eavesdropper of the key between users $i$ and $j$ makes estimate of $d_{ij}$ as:
\begin{align}
&\hat{d}_{ij}=\sqrt{\tilde{d}_{mi}^{2}+\tilde{d}_{mj}^{2}-2\tilde{d}_{mi}\tilde{d}_{mj}\cos(\tilde{\phi}_{m})}\label{estimatedij}
 \end{align}
where the parameters inside the square root are defined as \eqref{dij} and \eqref{phii}.
For $J\gg1$, $\sigma_{ij}^{2}/J\ll d_{ij}^{2}$ and $\sigma_{i}^{2}/J\approx 0$, $\forall i\neq j \in \{1,2,3\}$ with high probability and \eqref{estimatedij} can be approximated as \cite{mobility-conf}:
\begin{align}
\hat{d}_{ij}=d_{ij}+\mathcal{N}(0,\frac{\hat{\sigma}_{ij}^{2}}{J})\label{approxdij}
\end{align}
Substituting \eqref{approxdij} as the estimate of $d_{ij}$ in \eqref{apprate12}-\eqref{apprate23} results in the following rate region (it can be shown that this is the best that each potential eavesdropper can do):
\begin{theorem}\label{th2}
Using unlimited public channel in the pairwise key sharing from the localization information, all rates in the closure of the convex hull of the set of all key rate triples $(R_{12},R_{13},R_{23})$ that satisfy the following region, are achievable:
\begin{align}
&{R_{12} >0,R_{13} >0,R_{23} >0,}\no
\\&{R_{12}\leq \frac{1}{2}\mathbb{E}([\log(1+\frac{d_{12}^{4}J^{2}(\hat{\sigma}_{12}^{2}-\sigma_{12}^{2})}{(d_{12}^{2}J+\hat{\sigma}_{12}^{2})(2d_{12}^{2}J\sigma_{12}^{2}+\sigma_{12}^{4})})]^{+})}\no
\\&{R_{13}\leq \frac{1}{2}\mathbb{E}([\log(1+\frac{d_{13}^{4}J^{2}(\hat{\sigma}_{13}^{2}-\sigma_{13}^{2})}{(d_{13}^{2}J+\hat{\sigma}_{13}^{2})(2d_{13}^{2}J\sigma_{13}^{2}+\sigma_{13}^{4})})]^{+}))}\no
\\&{R_{23}\leq \frac{1}{2}\mathbb{E}([\log(1+\frac{d_{23}^{4}J^{2}(\hat{\sigma}_{23}^{2}-\sigma_{23}^{2})}{(d_{23}^{2}J+\hat{\sigma}_{23}^{2})(2d_{23}^{2}J\sigma_{23}^{2}+\sigma_{23}^{4})})]^{+})}\label{rrunlimitedpc}
\end{align}
in which $\mathbb{E}$ is the expectation with respect to $(d_{12},d_{13},d_{23})$ and
\begin{align}
\hat{\sigma}_{ij}^{2}\!\triangleq\!\sigma_{im}^{2}\!+\!\sigma_{jm}^{2}\!+ \! \text{Const}_{d_{12},d_{13},d_{23}}(\frac{\sigma_{m}^{2}}{4d_{ij}^{2}}\!-\!\frac{\sigma_{im}^{2}}{4d_{ij}^{2}d_{im}^{2}}\!-\!\frac{\sigma_{jm}^{2}}{4d_{ij}^{2}d_{jm}^{2}})\label{approxsigma}
\end{align}
for $\text{Const}_{d_{12},d_{13},d_{23}}=(d_{12}+d_{13}+d_{23})(d_{12}+d_{13}-d_{23})(d_{13}+d_{23}-d_{12})(d_{12}+d_{23}-d_{13})$.
\end{theorem}
\begin{IEEEproof}
The proof is given in Appendix B.
\end{IEEEproof}

In the following, we give an outer bound on the key capacity region in the described setup for unlimited public channel based on the explicit outer bound in \cite{salimi-pairwise}.

\begin{col}\label{col1}
Using unlimited public channel in the pairwise key agreement from localization information, the following is an outer bound on the pairwise key capacity region:
\begin{align}
&{R_{12} >0,R_{13} >0,R_{23} >0,}\no
\\&{R_{12}\leq \frac{1}{2}\log(1+\frac{\mathbb{E}(\hat{\sigma}_{12}^{2})}{\sigma_{12}^{2}})}\no
\\&{R_{13}\leq \frac{1}{2}\log(1+\frac{\mathbb{E}(\hat{\sigma}_{13}^{2})}{\sigma_{13}^{2}})}\no
\\&{R_{23}\leq \frac{1}{2}\log(1+\frac{\mathbb{E}(\hat{\sigma}_{23}^{2})}{\sigma_{23}^{2}})}
\end{align}
in which $\mathbb{E}$ is expected value with respect to $(d_{12},d_{13},d_{23})$ and $\hat{\sigma}_{ij}^{2}$ is defined as \eqref{approxsigma}.
\end{col}
\begin{IEEEproof}
The proof is given in Appendix C.
\end{IEEEproof}

\subsection{rate-limited public channel}\label{limitedpc}
In this case, the information sent by the users over the public channel should be subject to the respective rate constraints. In particular, a noisy version of the observation at each user can be considered for the key generation. To apply this constraint, we set:
\begin{align}
S_{ij}=\tilde{d}_{ij}+D_{ij}\label{noisyversion}
\end{align}
\noindent in Theorem 1 where $D_{ij}\!\sim\!{\rm{\mathcal N}}(0,\sigma'^{2}_{ij})$. The noises $D_{ij}$ are independent of each other and of all the observations. In fact $S_{ij}$ is a noisy version of $\tilde{d}_{ij}$ where its related information can be sent by user $i$ through the public channel with rate constraint $R_{i}$. It should be noted that in the case of rate-limited public channel, we can not assume one-way communication between each pair and we need to consider the general two-way communications to derive the largest rate region. By considering all the auxiliary random variables of Theorem 1 as \eqref{noisyversion} and applying the rate constraints in \eqref{cons1}-\eqref{cons123} in Theorem 1, we deduce:
\begin{theorem}\label{th3}
Using public channels with rates $(R_{1},R_{2},R_{3})$, respectively, at users 1,2,3 in the pairwise key sharing from localization information, the pairwise key rate region on the top of the next page is achievable which is subject to the constraints:
\small{
\begin{align}
&{\frac{1}{2}\mathbb{E}(\!\log(1\!+\!\frac{(2d_{12}^{2}J\!+\!\sigma_{12}^{2})\sigma_{12}^{2}}{(d_{12}^{2}J\!+\!\sigma_{12}^{2})\sigma_{12}^{'2}})\!+\!\log(1\!+\!\frac{(2d_{13}^{2}J\!+\!\sigma_{13}^{2})\sigma_{13}^{2}}{(d_{13}^{2}J\!+\!\sigma_{13}^{2})\sigma_{13}^{'2}}\!))\!\leq\! R_{1}}\no
\\&{\frac{1}{2}\mathbb{E}(\!\log(1\!+\!\frac{(2d_{12}^{2}J\!+\!\sigma_{12}^{2})\sigma_{12}^{2}}{(d_{12}^{2}J\!+\!\sigma_{12}^{2})\sigma_{21}^{'2}})\!+\!\log(1\!+\!\frac{(2d_{23}^{2}J\!+\!\sigma_{23}^{2})\sigma_{23}^{2}}{(d_{23}^{2}J\!+\!\sigma_{23}^{2})\sigma_{23}^{'2}}\!))\!\leq\! R_{2}}\no
\\&{\frac{1}{2}\mathbb{E}(\!\log(1\!+\!\frac{(2d_{13}^{2}J\!+\!\sigma_{13}^{2})\sigma_{13}^{2}}{(d_{13}^{2}J\!+\!\sigma_{13}^{2})\sigma_{31}^{'2}})\!+\!\log(1\!+\!\frac{(2d_{23}^{2}J\!+\!\sigma_{23}^{2})\sigma_{23}^{2}}{(d_{23}^{2}J\!+\!\sigma_{23}^{2})\sigma_{32}^{'2}}\!))\!\leq\! R_{3}}\label{noisyconstraints}
\end{align}}

\begin{figure*}[!t]
\tiny{
\begin{align}
&{R_{12} >0,R_{13} >0,R_{23} >0,}\no
\\ \hspace{-3cm}&{R_{12}\leq \frac{1}{2}\mathbb{E}([\log(1+\frac{d_{12}^{4}J^{2}(\hat{\sigma}_{12}^{2}-\sigma_{12}^{2})}{(d_{12}^{2}J+\hat{\sigma}_{12}^{2})(d_{12}^{2}J(2\sigma_{12}^{2}+\sigma_{12}^{'2})+(\sigma_{12}^{2}+\sigma_{12}^{'2})\sigma_{12}^{2})})]^{+}+[\log(1+\frac{d_{12}^{4}J^{2}(\hat{\sigma}_{12}^{2}\sigma_{12}^{'2}-\sigma_{12}^{2}(\sigma_{12}^{2}+\sigma_{12}^{'2}))}{(d_{12}^{2}J(\hat{\sigma}_{12}^{2}+\sigma_{12}^{2}+\sigma_{12}^{'2})+\hat{\sigma}_{12}^{2}(\sigma_{12}^{2}+\sigma_{12}^{'2}))(d_{12}^{2}J(2\sigma_{12}^{2}+\sigma_{21}^{'2})+(\sigma_{12}^{2}+\sigma_{21}^{'2})\sigma_{12}^{2})})]^{+})}\no
\\&{R_{13}\leq \frac{1}{2}\mathbb{E}([\log(1+\frac{d_{13}^{4}J^{2}(\hat{\sigma}_{13}^{2}-\sigma_{13}^{2})}{(d_{13}^{2}J+\hat{\sigma}_{13}^{2})(d_{13}^{2}J(2\sigma_{13}^{2}+\sigma_{13}^{'2})+(\sigma_{13}^{2}+\sigma_{13}^{'2})\sigma_{13}^{2})})]^{+}+[\log(1+\frac{d_{13}^{4}J^{2}(\hat{\sigma}_{13}^{2}\sigma_{13}^{'2}-\sigma_{13}^{2}(\sigma_{13}^{2}+\sigma_{13}^{'2}))}{(d_{13}^{2}J(\hat{\sigma}_{13}^{2}+\sigma_{13}^{2}+\sigma_{13}^{'2})+\hat{\sigma}_{13}^{2}(\sigma_{13}^{2}+\sigma_{13}^{'2}))(d_{13}^{2}J(2\sigma_{13}^{2}+\sigma_{32}^{'2})+(\sigma_{13}^{2}+\sigma_{31}^{'2})\sigma_{13}^{2})})]^{+})}\no
\\&{R_{23}\leq \frac{1}{2}\mathbb{E}([\log(1+\frac{d_{23}^{4}J^{2}(\hat{\sigma}_{23}^{2}-\sigma_{23}^{2})}{(d_{23}^{2}J+\hat{\sigma}_{23}^{2})(d_{23}^{2}J(2\sigma_{23}^{2}+\sigma_{23}^{'2})+(\sigma_{23}^{2}+\sigma_{23}^{'2})\sigma_{23}^{2})})]^{+}[\log(1+\frac{d_{23}^{4}J^{2}(\hat{\sigma}_{23}^{2}\sigma_{23}^{'2}-\sigma_{23}^{2}(\sigma_{23}^{2}+\sigma_{23}^{'2}))}{(d_{23}^{2}J(\hat{\sigma}_{23}^{2}+\sigma_{23}^{2}+\sigma_{23}^{'2})+\hat{\sigma}_{23}^{2}(\sigma_{23}^{2}+\sigma_{23}^{'2}))(d_{23}^{2}J(2\sigma_{23}^{2}+\sigma_{32}^{'2})+(\sigma_{23}^{2}+\sigma_{32}^{'2})\sigma_{23}^{2})})]^{+})}\label{rrunlimitedpc}
\end{align}}
\hrulefill
\end{figure*}
\end{theorem}
\normalsize{
\begin{IEEEproof}
The proof is given in Appendix B.
\end{IEEEproof}}

\section{Numerical Results}
In this section, numerical evaluation of the results in Sections IV-A and IV-B is given. We assume that at each time slot, all users' locations are characterized by i.i.d. circularly symmetric zero mean, unit variance Gaussian random variables. First we consider unlimited public channel case. We set $\sigma_{13}^{2}=\sigma_{23}^{2}=\sigma_{1}^{2}=\sigma_{2}^{2}=\sigma_{3}^{2}=0.1$ and plot the key rates as functions of $\sigma_{12}^{2}$. Because of symmetry, the bounds on the rates $R_{13}$ and $R_{23}$ are the same and hence, we analyse one of them. In Fig. 3, the inner and outer bounds on key rates $R_{12}$ and $R_{13}$ are shown as functions of $\sigma_{12}^{2}$. Clearly the bounds on $R_{12}$ decrease as $\sigma_{12}^{2}$ increases, while the bounds on $R_{13}$ increase with the growth of $\sigma_{12}^{2}$. However, for small values of $\sigma_{12}^{2}$, the bounds on $R_{12}$ are more affected compared to the bounds on $R_{13}$.
\begin{figure}%[!t]
\centering
%\vspace{-2.5cm}
\includegraphics[width=8cm]{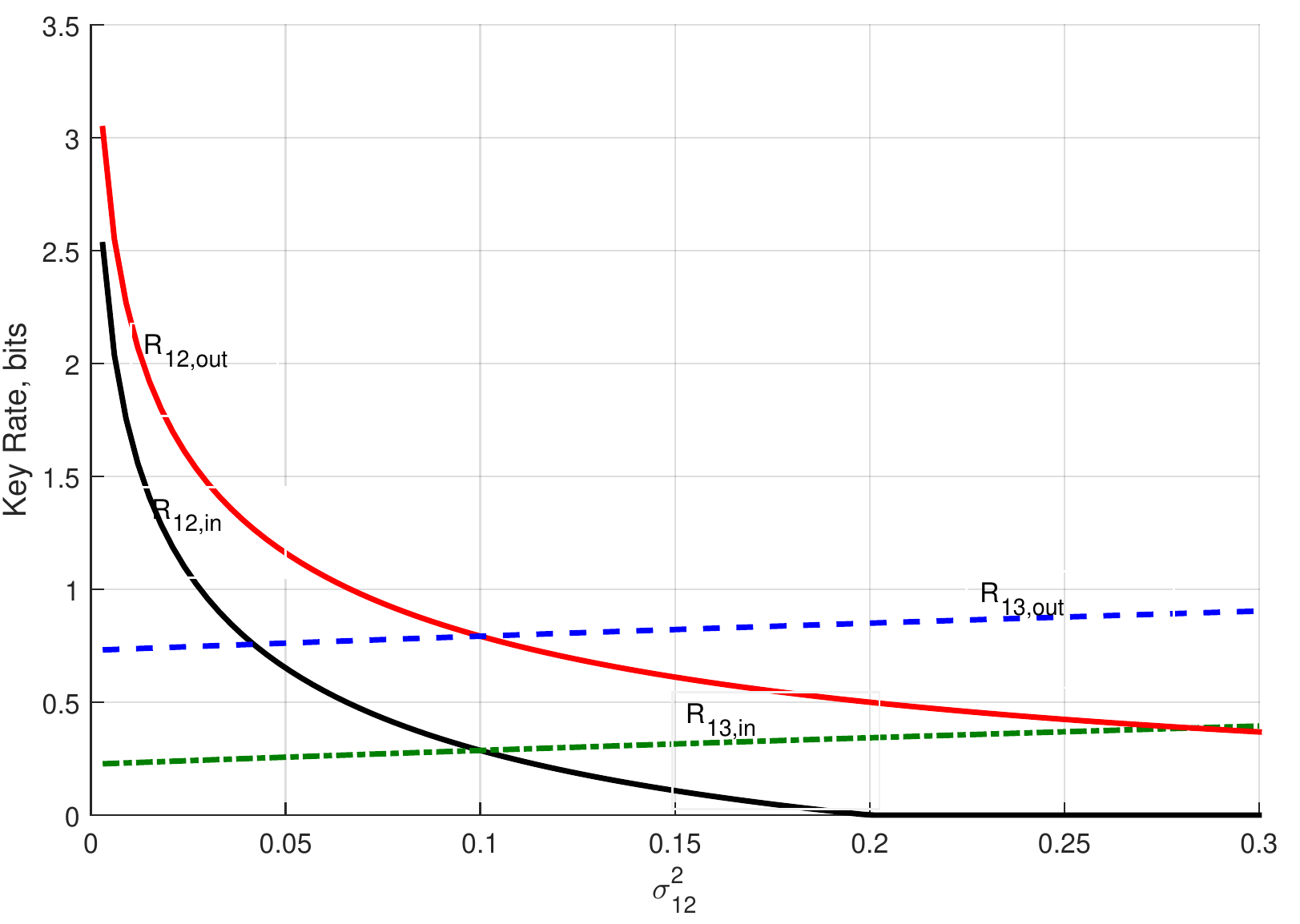}
%\vspace{-2.5cm}
\footnotesize{
\caption{inner and outer bounds on $R_{12}$ and $R_{13}$}
}
\vspace{-.2cm}
\label{fig_sim}
\end{figure}

%\begin{figure}%[!t]
%\centering
%\vspace{-.5cm}
%\includegraphics[width=9cm]{unlimited.pdf}
%\vspace{-.5cm}
%\footnotesize{
%\caption{inner and outer bounds on $R_{12}$ and $R_{13}$}
%}
%\end{figure}

Then, we analyse the key rate region in the rate-limited public channel case. We set $R_{1}=.5$, $R_{2}=.2$, $R_{3}=.8$ and $\sigma_{12}^{2}=\sigma_{13}^{2}=\sigma_{23}^{2}=\sigma_{1}^{2}=\sigma_{2}^{2}=\sigma_{3}^{2}=0.1$. In order to clarify the rate region, we project the 3-D region into three 2-D regions. As we discussed in Section IV-B, in the case of rate-limited pubic channel, we have two-way communication between each pair. Each user splits its available public channel rate to share keys with the other users while the public channel rates of the other users affect this splitting. As shown in Fig. 4--6, the rate regions are not necessarily rectangular in contrast to the case of unlimited public channel. Obviously, the achievable rates are significantly smaller than the corresponding values in Fig. 3 where unlimited public channel is assumed (respective rates at Fig. 3 for $\sigma_{12}^{2}=0.1$).

%\begin{figure*}[!t]
%    \fbox{\includegraphics[width=5.5cm]{R12+13.pdf}}
%    \hspace{10px}
%    \fbox{\includegraphics[width=5.5cm]{R12+23.pdf}}
%    \hspace{10px}
%    \fbox{\includegraphics[width=5.5cm]{R13+23.pdf}}
%    \caption{Projection of the rate region into three 2-D rate regions for rate-limited public channels with $R_{1}=.5$, $R_{2}=.2$,$R_{3}=.8$}
%\end{figure*}
\begin{figure}
\vspace{-.2cm}
    \includegraphics[width=7cm]{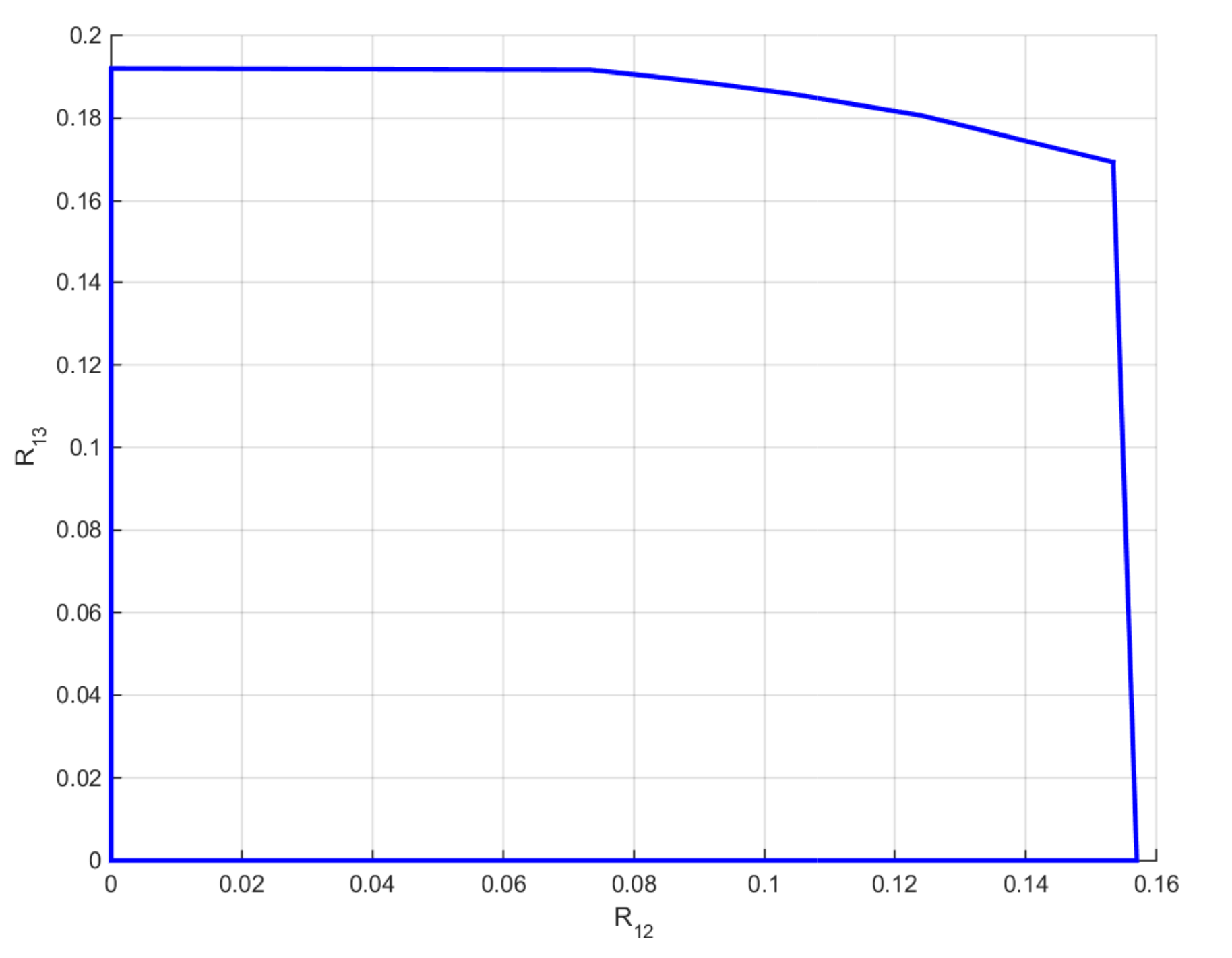}
\vspace{-.2cm}
    \caption{$R_{12}-R_{13}$ with $R_{1}=.5$, $R_{2}=.2$,$R_{3}=.8$}
    \vspace{-.2cm}
\end{figure}

 \begin{figure}
    \includegraphics[width=7cm]{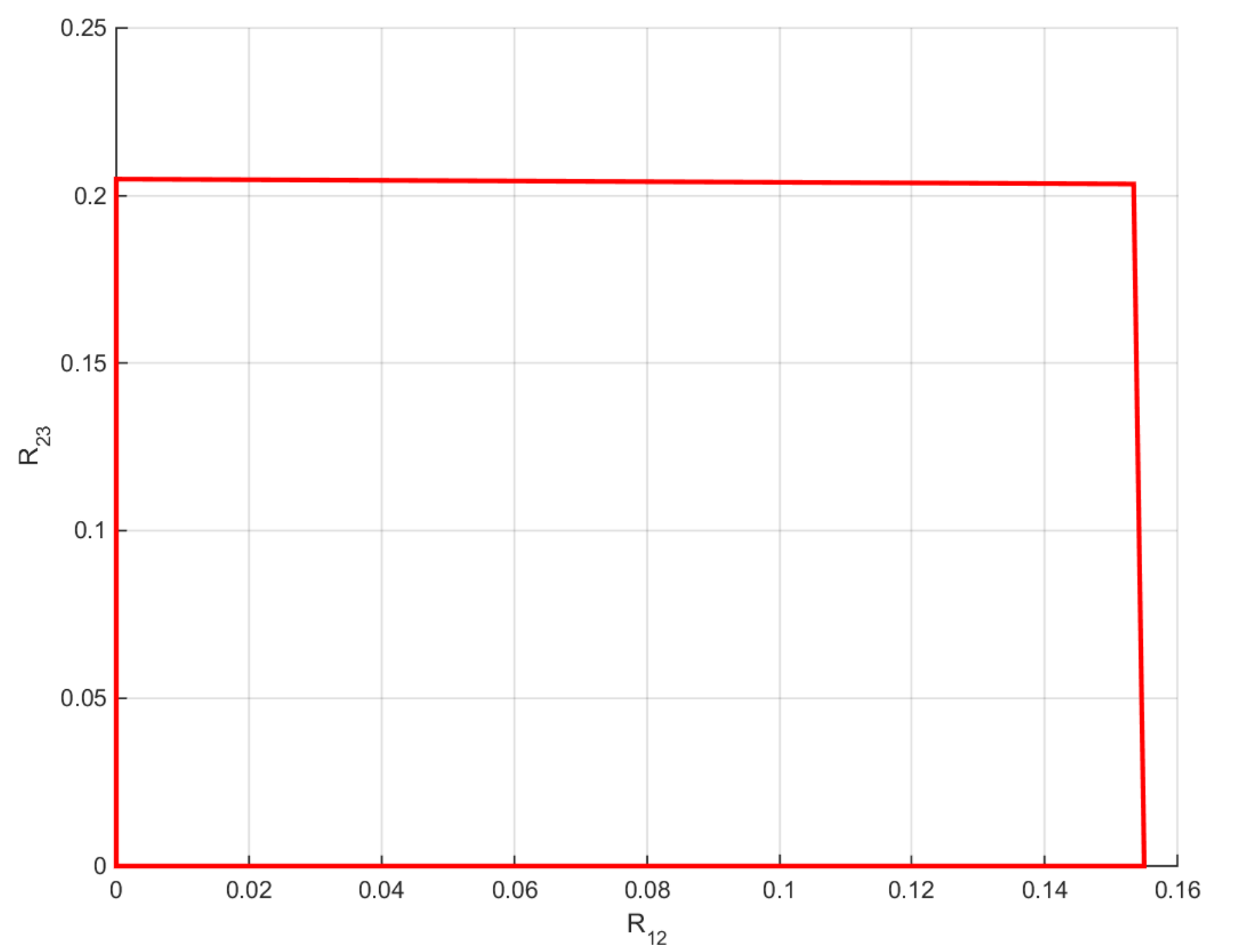}
    \vspace{-.2cm}
    \caption{$R_{12}-R_{23}$ with $R_{1}=.5$, $R_{2}=.2$,$R_{3}=.8$}
    \vspace{-.2cm}
\end{figure}
\begin{figure}
\vspace{-.2cm}
   \includegraphics[width=7cm]{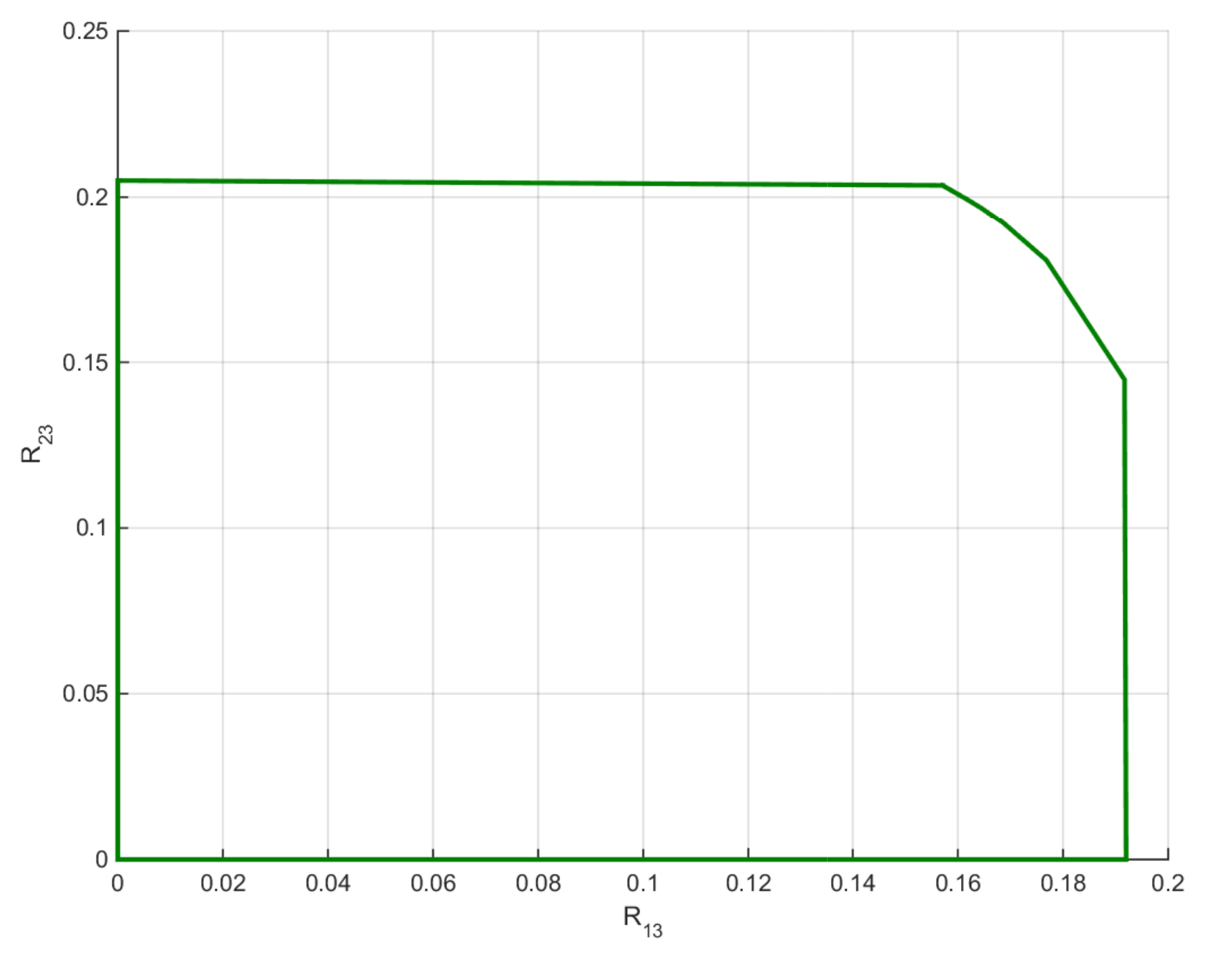}
    \vspace{-.2cm}
    \caption{$R_{13}-R_{23}$ with $R_{1}=.5$, $R_{2}=.2$,$R_{3}=.8$}
    \vspace{-.2cm}
\end{figure}
\section{Conclusion}
\noindent
The source model of pairwise secret key sharing was investigated with rate-limited pubic channel between three users.
An inner bound on the key capacity region was derived for the general case of discrete memoryless source observations. We considered a setup in which the users exploited the distance between themselves as correlated observations to generate keys. Inner and outer bounds on the key capacity region were analyzed for the case of i.i.d. Gaussian observations. As a future work, we analyze the problem of pairwise key sharing between arbitrary number of users who access to limited public channel.

\section*{Appendix A}

\noindent
\subsection*{Proof of Theorem 1}

\noindent We fix the distribution the same as in Theorem 1. Double random binning, rate splitting technique and Wyner-Ziv coding are used to prove the achievability of the region presented in Theorem 1. The total key between each pair of the users consists of two parts; each part is generated by one of them. In continue, we describe code construction, encoding, decoding and security analysis. In the following, a random variable is denoted
by an upper case letter and its realization is denoted by the corresponding lower case letter. $\textbf{X}$ (resp. $\textbf{x}$). denotes $n$ repetitions of random variable $X$, i.e., $X^{n}$ (resp. its realization $x^{n}$).

\indent
$S_{ij}$ denotes the auxiliary random variable associated with key $K_{ij}$ generated by user $i$ to be shared with user $j$. For code construction, user $1$ chooses $2^{n(\!r_{12}\!+\!r'_{12}\!+\!r''_{12}\!)}$ and $2^{n(\!r_{13}\!+\!r'_{13}\!+\!r''_{13}\!)}$ sequences $\textbf{s}_{12}$ and $\textbf{s}_{13}$ from $A_{\varepsilon'}^{n}(P_{S_{12}})$ and $A_{\varepsilon'}^{n}(P_{S_{13}})$, respectively, in which $\varepsilon'>0$ can be chosen arbitrarily small. $A_{\varepsilon'}^{n}(\!P_{X}\!)$ denotes a set of $\varepsilon'-$typical
sequences $x^{n}$ with respect to distribution $p(\!x)$. These sequences are labeled as $\textbf{s}_{12}(k_{12},\!k'_{12},\!k''_{12})$ and $\textbf{s}_{13}(k_{13},\!k'_{13},\!k''_{13})$ such that:
\small{
\[\begin{array}{l}
\hspace{-.2cm}{\!k_{12}\!\!\in \!\!\mathcal K_{12}\!=\!\!\{\!1,\!...,\!2^{nr_{12}}\!\}\!,\!k'_{12}\!\!\in\!\!\mathcal K'_{12}\!=\!\!\{\!1,\!...,\!2^{nr'_{12}}\!\},\!k''_{12}\!\!\in\!\!\mathcal K''_{12}\!\!=\!\!\{\!1,\!...,\!2^{nr''_{12}}\!\},\!}
\\ \hspace{-.2cm}{\!k_{13}\!\!\in \!\!\mathcal K_{13}\!=\!\!\{\!1,\!...,\!2^{nr_{13}}\!\}\!,\!k'_{13}\!\!\in\!\!\mathcal K'_{13}\!=\!\!\{\!1,\!...,\!2^{nr'_{13}}\!\},\!k''_{13}\!\!\in\!\!\mathcal K''_{13}\!=\!\!\{\!1,\!...,\!2^{nr''_{13}}\!\},\!}
\end{array}\]}

\normalsize{
Sequences $\textbf{s}_{12}$ and $\textbf{s}_{13}$ are produced by user 1 to share secret keys with user 2 and user 3, respectively. Similarly, sequences $\textbf{s}_{21}(k_{21},\!k'_{21},\!k''_{21})$ and $\textbf{s}_{23}(k_{23},\!k'_{23},\!k''_{23})$ are generated by user 2 to share secret keys with users 1 and 3, respectively, and sequences $\textbf{s}_{31}(k_{31},\!k'_{31},\!k''_{31})$ and $\textbf{s}_{32}(k_{32},\!k'_{32},\!k''_{32})$ are generated by user 3 to share secret keys with users 1 and 2, respectively.
These sequences are similarly labeled using double random binning. We choose:}
\begin{align}
&{r''_{12}\!+r''_{21}=\!I(S_{12},S_{21};X_{3},S_{13},S_{23})\!+\!I(S_{12};S_{21})-\varepsilon'}\label{r12+r21}
\\&{r''_{13}\!+r''_{31}=\!I(S_{13},S_{31};X_{2},S_{12},S_{32})\!+\!I(S_{13};S_{31})-\varepsilon'}\label{r13+r31}
\\&{r''_{23}\!+r''_{32}=\!I(S_{23},S_{32};X_{1},S_{31},S_{21})\!+\!I(S_{23};S_{32})-\varepsilon'}\label{r23+r32}
\end{align}
\noindent
\indent
For encoding, user 1 looks for sequences $\textbf{s}_{12}$ and $\textbf{s}_{13}$; each is $\varepsilon'-$jointly typical with $\textbf{x}_{1}$ and declares error if there are no such sequences. Symmetrically, $(\textbf{s}_{21},\textbf{s}_{23})$ and $(\textbf{s}_{31},\textbf{s}_{32})$ are respectively chosen by user 2 and user 3 based on their observations $\textbf{x}_{2}$ and $\textbf{x}_{3}$. According to Covering Lemma \cite{elgamalbook}, the error probability of choosing these sequences would be arbitrarily small if we have:
\begin{align}
&{r_{12}+r'_{12}+r''_{12}\!=\!I(S_{12};X_{1})+\varepsilon'',}\label{typicality12}
\\&{r_{13}+r'_{13}+r''_{13}\!=\!I(S_{13};X_{1})+\varepsilon'',}\label{typicality13}
\\&{r_{21}+r'_{21}+r''_{21}=\!I(S_{21};X_{2})+\varepsilon'',}\label{typicality21}
\\&{r_{23}+r'_{23}+r''_{23}\!=\!I(S_{23};X_{2})+\varepsilon'',}\label{typicality23}
\\&{r_{31}+r'_{31}+r''_{31}=\!I(S_{31};X_{3})+\varepsilon''.}\label{typicality31}
\\&{r_{32}+r'_{32}+r''_{32}\!=\!I(S_{32};X_{3})+\varepsilon''.}\label{typicality32}
\end{align}
\noindent in which $\varepsilon''>0$ is chosen such that $\varepsilon''>\varepsilon'$, (e.g., $\varepsilon''=2\varepsilon'$). Not that there is no need to take into account the joint typicality of $\textbf{s}_{12}$ and $\textbf{s}_{13}$ separately since according to the distribution of random variables in Theorem 1, $I(S_{12},S_{13}|X_{1})=0$. The same holds for the auxiliary random variables at users 2 and 3. Then, user $i$ selects the respective index $k_{ij}$ of $\textbf{s}_{ij}$ as the key for sharing with user $j$. For such sequence, the respective index $k'_{ij}$ is the required information to be sent from user $i$ to user $j$ such that user $j$ can decode the corresponding key. Thus, user 1 sends $k'_{12}$ and $k'_{13}$ to users 2 and 3, respectively, through its public channel with rate constraint $R_{1}$. Similarly users 2 sends $k'_{21}$ and $k'_{23}$ and 3 sends $k'_{31}$ and $k'_{32}$ to the respective users through their public channels with rate constraints $R_{2}$ and $R_{3}$. We assume:
\begin{align}
&{r'_{12}+r'_{13}\leq R_{1},}\label{user1limit}
\\&{r'_{21}+r'_{23}\leq R_{2},}\label{user2limit}
\\&{r'_{31}+r'_{32}\leq R_{3},}\label{user3limit}
\end{align}
Next in the decoding step, it is shown that \eqref{user1limit}-\eqref{user3limit} hold according to the rate constraints \eqref{cons1}-\eqref{cons123} in Theorem 1.

\indent For decoding, according to \eqref{user1limit}-\eqref{user3limit}, we assume that user 1 has received $k'_{21}$ and $k'_{31}$ from users 2 and 3, respectively with arbitrarily small probability of error. With access to observation $\textbf{x}_{1}$ and indices $k'_{21}$ and $k'_{31}$, user 1 chooses sequences $\textbf{s}_{21}$ and $\textbf{s}_{31}$ such that:%which are $\varepsilon _{1}-$jointly typical with $\textbf{x}_{1}$ where $\varepsilon_{1}=\frac{\varepsilon}{4}$.
%Let $A_{\varepsilon_{1}}^{N}(P_{S_{21},S_{31},X_{1}})$ denote a set of $\epsilon_{1}-$jointly typical sequences $(s_{21}^{N},s_{31}^{N},x_{1}^{N})$ with respect to the distribution $p_{s_{21},s_{31},x_{1}}$. User 1 decodes key pair $(k_{21},k_{31})$ if:
\begin{equation}
(\textbf{s}_{21}(k_{21},k'_{21},k''_{21}),\textbf{s}_{31}(k_{31},k'_{31},k''_{31}),\textbf{x}_{1})\in A_{\varepsilon_{1}}^{n}(P_{S_{21},S_{31},X_{1}}),\no
\end{equation}
when such $(\textbf{s}_{21},\textbf{s}_{31})$ exists and is unique. Otherwise, it declares error. Symmetrically, users 2 and 3 decode the sequence pairs $(\textbf{s}_{12}(k_{12},k'_{12},k''_{12}),s_{32}(k_{32},k'_{32},k''_{32}))$ and $(\textbf{s}_{13}(k_{13},k'_{13},k''_{13}),s_{23}(k_{23},k'_{23},k''_{23}))$, respectively.
It can be shown that the decoding error probability at the users is bounded as ($P_{ei}^{(n)}$is the decoding error probability at user $i$):
\[\begin{array}{l} {P_{e1}^{(n)} \le \varepsilon _{1} +2^{n(4\varepsilon _{1} +I(S_{21},S_{31};X_{2},X_{3}|X_{1})-(r'_{21} +r'_{31} ))} +} \\ {{\rm \; \;}2^{n(3\varepsilon _{1} +I(S_{21};X_{2}|S_{31} ,X_{1})-r'_{21})}+2^{n(3\varepsilon _{1}+I(S_{31};X_{3}|S_{21},X_{1})-r'_{31} )},}
\\ {P_{e2}^{(n)} \le \varepsilon _{1} +2^{n(4\varepsilon _{1} +I(S_{12},S_{32};X_{1},X_{3}|X_{2})-(r'_{12} +r'_{32} ))} +} \\ {{\rm \; \;}2^{n(3\varepsilon _{1} +I(S_{12};X_{1}|S_{32} ,X_{2})-r'_{12})}+2^{n(3\varepsilon _{1}+I(S_{32};X_{3}|S_{12},X_{2})-r'_{32} )},}
\\{P_{e3}^{(n)} \le \varepsilon _{1} +2^{n(4\varepsilon _{1} +I(S_{13},S_{23};X_{1},X_{2}|X_{3})-(r'_{13} +r'_{23} ))} +} \\ {{\rm \; \;}2^{n(3\varepsilon _{1} +I(S_{13};X_{1}|S_{23} ,X_{3})-r'_{13})}+2^{n(3\varepsilon _{1}+I(S_{23};X_{2}|S_{13},X_{3})-r'_{23} )}.}

\end{array}\]
If we set:\small{
\begin{align}
&{r'_{21} +r'_{31} >I(S_{21} ,S_{31};X_{2},X_{3}|X_{1})+2\varepsilon'}\label{sent1}
\\&{r'_{21}>I(S_{21};X_{2}|S_{31},X_{1})+\varepsilon'}
\\&{r'_{31}>I(S_{31};X_{3}|S_{21},X_{1})+\varepsilon'}
\\&{r'_{12} +r'_{32} >I(S_{12} ,S_{32};X_{1},X_{3}|X_{2})+2\varepsilon'}
\\&{r'_{12}>I(S_{12};X_{1}|S_{32},X_{2})+\varepsilon'}
\\&{r'_{32}>I(S_{32};X_{3}|S_{12},X_{2})+\varepsilon'}
\\&{r'_{13} +r'_{23} >I(S_{13} ,S_{23};X_{1},X_{2}|X_{3})+2\varepsilon'}
\\&{r'_{13}>I(S_{13};X_{1}|S_{23},X_{3})+\varepsilon'}
\\&{r'_{23}>I(S_{23};X_{2}|S_{13},X_{3})+\varepsilon'}\label{sent2}
\end{align}
}
\normalsize{\noindent then for $i=1,2,3$ we have:
\noindent\begin{equation}
P_{ei}^{(n)}\le\varepsilon_{1} +2^{n(4\varepsilon_{1} -2\varepsilon')}+2^{n(3\varepsilon_{1} -\varepsilon')} +2^{n(3\varepsilon_{1}-\varepsilon')} . \no
\end{equation}
By setting $\varepsilon_{1}\!=\!\frac{\varepsilon}{32}$ and $\varepsilon'\!=4\varepsilon_{1}=\frac{\varepsilon}{8}$, we choose $n$ sufficiently large that $2^{-n\varepsilon_{1}}\!\le\!\varepsilon_{1}$, and then $P_{ei}^{(n)}\!\le\!4\varepsilon _{1}\!=\!\frac{\varepsilon}{8}$.

After these steps $K_{i,j}=(K_{ij},K_{ji})$ is shared between users $i$ and $j$.
}
%Replacing equations \eqref{GrindEQ__13_}-\eqref{GrindEQ__21_} in \eqref{GrindEQ__10_}-\eqref{GrindEQ__12_}, we obtain:
\noindent Replacing equations \eqref{sent1}-\eqref{sent2} and \eqref{r12+r21}-\eqref{r23+r32} in \eqref{typicality12}-\eqref{typicality32}, we obtain:
\small{
\[\begin{array}{l} {r_{12} +r_{21} \le {\bf r}_{{\bf 12}} {\bf +r}_{{\bf 21}} -{\bf I}_{{\bf 12}},} \\ {r_{13} +r_{31} \le {\bf r}_{{\bf 13}} {\bf +r}_{{\bf 31}} -{\bf I}_{{\bf 13}},}
\\ {r_{23} +r_{32} \le {\bf r}_{{\bf 23}} {\bf +r}_{{\bf 32}} -{\bf I}_{{\bf 23}},}
\\ {r_{12} +r_{21} +r_{13} +r_{31} \le {\bf r}_{{\bf 12}} {\bf +r}_{{\bf 21}}{\bf +r}_{{\bf 13}}{\bf +r}_{{\bf 31}} -{\bf I}_{{\bf 12}}-{\bf I}_{{\bf 13}}-{\bf I}_{{\bf 1}},}
\\ {r_{12} +r_{21} +r_{23} +r_{32} \le {\bf r}_{{\bf 12}} {\bf +r}_{{\bf 21}} {\bf +r}_{{\bf 23}} {\bf +r}_{{\bf 32}} -{\bf I}_{{\bf 12}}-{\bf I}_{{\bf 23}}-{\bf I}_{{\bf 2}},}
\\ {r_{13} +r_{31} +r_{23} +r_{32} \le {\bf r}_{{\bf 13}} {\bf +r}_{{\bf 31}}{\bf +r}_{{\bf 23}} {\bf +r}_{{\bf 32}} -{\bf I}_{{\bf 13}}-{\bf I}_{{\bf 23}}-{\bf I}_{{\bf 3}},}
\\ {r_{12} +r_{21} +r_{13} +r_{31} +r_{23} +r_{32} \le {\bf r}_{{\bf 12}} {\bf +r}_{{\bf 21}} {\bf +r}_{{\bf 13}} {\bf +r}_{{\bf 31}} {\bf +r}_{{\bf 23}}+ } \\ {{\bf r}_{{\bf 32}}-{\bf I}_{{\bf 12}} -{\bf I}_{{\bf 13}}-{\bf I}_{{\bf 23}}-{\bf I}_{{\bf 1}}-{\bf I}_{{\bf 2}}-{\bf I}_{{\bf 3}}.} \end{array}\]
}\normalsize{\noindent By setting $R_{12}=r_{12}+r_{21},R_{13}=r_{13}+r_{31},R_{23}=r_{23}+r_{32},$ and applying Fourier-Motzkin elimination \cite{fouri-motz} to the above region, the rate region of Theorem 1 can be derived.
}

\begin{rem}
The necessary conditions \eqref{user1limit}-\eqref{user3limit} hold according to the rate constraints \eqref{cons1}-\eqref{cons123} in Theorem 1 and equations \eqref{sent1}-\eqref{sent2}.
\end{rem}

\indent Now, we should check the security conditions of definition 1. We give the proof of \eqref{eq3} for $i=1$, $j=2$, and $m=3$. By symmetry, the other security conditions are deduced. By substituting $K_{1,2}=(K_{12},K_{21})$, $F_{1}=(K'_{12},K'_{13})$ and $F_{2}=(K'_{21},K'_{23})$ we have:
\vspace{.2cm}
\small{

\noindent
$\begin{array}{l}{I(K_{12},K_{21} ;\textbf{X}_{3},K'_{12},K'_{13},K'_{21},K'_{23})}
\\ {\mathop{\le}\limits^{(a)}I(K_{12},K_{21};\textbf{X}_{3},\textbf{S}_{13},\textbf{S}_{23},K'_{12},K'_{21})}
\vspace{.2cm}
\\ {=H(K_{12},K_{21})-H(K_{12},K_{21}|\textbf{X}_{3},\textbf{S}_{13},\textbf{S}_{23},K'_{12},K'_{21})}
\\ {=H(K_{12},K_{21})-H(K_{12},K_{21},K'_{12},K'_{21}|\textbf{X}_{3},\textbf{S}_{13},\textbf{S}_{23})}
\\{+H(K'_{12},K'_{21}|\textbf{X}_{3},\textbf{S}_{13},\textbf{S}_{23})}
\vspace{.1cm}
\\ {\leq H(\!K_{12},\!K_{21}\!)\!-\!H(K_{12},\!K_{21},\!K'_{12},\!K'_{21}|\textbf{X}_{3},\!\textbf{S}_{13},\!\textbf{S}_{23}\!)\!+\!H(\!K'_{12},K'_{21}\!)}
\\ {= H(K_{12},K_{21})-H(K_{12},K_{21},K'_{12},K'_{21},K''_{12},K''_{21}|\textbf{X}_{3},\textbf{S}_{13},\textbf{S}_{23})}
\\ {+H(K'_{12},K'_{21})+ H(K''_{12},K''_{21}|K_{12},K_{21},K'_{12},K'_{21},\textbf{X}_{3},\textbf{S}_{13},\textbf{S}_{23})}
\\ {\mathop{\le}\limits^{(b)} H(K_{12},K_{21})-H(K_{12},K_{21},K'_{12},K'_{21},K''_{12},K''_{21}|\textbf{X}_{3},\textbf{S}_{13},\textbf{S}_{23})}
\\{+H(K'_{12},K'_{21})+n\varepsilon _{2}}
\\ {\mathop{=}\limits^{(c)} H(K_{12},K_{21})-H(\textbf{S}_{12},\textbf{S}_{21}|\textbf{X}_{3},\textbf{S}_{13},\textbf{S}_{23})\!+\!H(K'_{12},K'_{21})\!+\!n\varepsilon _{2}}
\\{\mathop{\le}\limits^{(d)}\!\!H(K_{12},K_{21})\!+\!H(K'_{12},K'_{21})\!-nH(S_{12},S_{21}|X_{3},S_{13},S_{23})}
\\{+n(\varepsilon_{2}+\varepsilon_{3})}
\vspace{.1cm}
\\{=\!\!H(K_{12})+H(K_{21})\!+\!H(K'_{12},K'_{21})\!}
\\{-nH(S_{12},S_{21}|X_{3},S_{13},S_{23})+n(\varepsilon_{2}+\varepsilon_{3})-I(K_{12};K_{21})}
\vspace{.1cm}
\\{\leq\!\!H(K_{12})+H(K_{21})\!+\!H(K'_{12})+H(K'_{21})\!}
\\{-nH(S_{12},S_{21}|X_{3},S_{13},S_{23})+n(\varepsilon_{2}+\varepsilon_{3})-I(K_{12};K_{21})}
\vspace{.1cm}
\\{=\!\!(\!H(K_{12}\!)+H(K_{21}\!)\!+\!H(K'_{12})\!+\!H(K'_{21})+\!H(\!K''_{12})+H(K''_{21})\!)}
\\{\!-\!H(\!K''_{12}\!)\!\!-\!\!H(\!K''_{21}\!)\!-\!n(\!H(\!S_{\!12},\!S_{21}|X_{3},\!S_{\!13},\!S_{23}\!)\!-\!\varepsilon_{2}\!\!-\!\!\varepsilon_{3}\!)\!-\!I(\!K_{\!12};\!K_{21}\!)}
\\{\mathop{=}\limits^{(e)}-nH(\!S_{\!12}|X_{1}\!)\!-\!nH(\!S_{21}|X_{2}\!)\!-\!I(\!K_{12};K_{21}\!)\!\!+\!n(\!2\varepsilon''\!+\!\varepsilon'\!+\!\varepsilon_{2}\!+\!\varepsilon_{3}\!)}
\\{\leq n(2\varepsilon''+\varepsilon'+\varepsilon_{2}+\varepsilon_{3})-I(K_{12};K_{21})}
\\{\leq n(2\varepsilon''+\varepsilon'+\varepsilon_{2}+\varepsilon_{3})=n(5\varepsilon'+\varepsilon_{2}+\varepsilon_{3})}
\end{array}$
}

\normalsize{
\noindent In the above equations, (a) follows from the fact that $k'_{13}$ and $k'_{23}$ are induces of the sequences $s_{13}^{N}$ and $s_{23}^{N}$. To prove (b), the same approach as lemma 2 in \cite{salimi} can be exploited to show $H(\textbf{S}_{12},\textbf{S}_{21}|\textbf{X}_{3},\textbf{S}_{13},\textbf{S}_{23},K'_{12},K'_{21}\,K_{12},K_{21})\le n\varepsilon_{2}$ (based on the rates defined in \eqref{r12+r21}-\eqref{r23+r32}). (c) is due to the fact that that with access to $(k_{12},k'_{12},k''_{12})$ and $(k_{21},k'_{21},k''_{21})$ sequences $\textbf{s}_{12}$ and $\textbf{s}_{21}$ are determined. To prove (d), the same approach as lemma 1 in \cite{salimi} can be exploited to show $H(\textbf{S}_{12},\textbf{S}_{21}|\textbf{X}_{3},\textbf{S}_{13},\textbf{S}_{23})\ge nH(S_{12},S_{21}|X_{3},S_{13},S_{23} )-n\varepsilon_{3}$. (e) is followed from the definition of the rates in \eqref{r12+r21}, \eqref{typicality12} and \eqref{typicality21}.
\normalsize{
\noindent By defining $\varepsilon_{2}=\varepsilon_{3}=\frac{\varepsilon}{8}$, we obtain:
\begin{equation}\no
I(K_{12},K_{21} ;\textbf{X}_{3},K'_{12},K'_{13},K'_{21},K'_{23})\leq n(5\varepsilon'+\varepsilon_{2}+\varepsilon_{3})=n\frac{7\varepsilon}{8}
\end{equation}

To show that the total rate of the key between users 1 and 2 is the sum of rates $r_{12}$ and $r_{21}$, we should prove the independence
of $K_{12}$ and $K_{21}$. When analyzing the security condition, we showed that:
\small{
\begin{equation}\no
I(\!K_{12},K_{21} ;\textbf{X}_{3},K'_{12},K'_{13},K'_{21},K'_{23}\!)\leq n(\!5\varepsilon'+\varepsilon_{2}+\varepsilon_{3}\!)\!-\!I(\!K_{12};K_{21}\!)
\end{equation}}
\normalsize{
which implies:}
\begin{equation}\no
I(K_{12};K_{21})\leq n(5\varepsilon'+\varepsilon_{2}+\varepsilon_{3})
\end{equation}
and hence, we deduce the independence of the keys.

\section*{Appendix B}
\noindent
\subsection*{Proof of Theorem 2}
\noindent In the case of unlimited public channel, it is assumed that user 1 communicates to user 2, user 2 communicates to user 3 and user 3 communicates to user 1.
According to the directions of communications between users, we choose $S_{12}=\tilde{d}_{12},S_{23}=\tilde{d}_{23},S_{31}=\tilde{d}_{31},S_{21}=S_{32}=S_{13}=null$ in Theorem 1. Then the rate region in Theorem 1 is reduced to:
\begin{align}
&{R_{12} >0,R_{13} >0,R_{23} >0,}
\\&{R_{12}\leq I\!(\!\tilde{d}_{12};\!\tilde{d}_{\!21}\!)\!-\!\!I\!(\!\tilde{d}_{12};\tilde{d}_{\!31},\tilde{d}_{\!32},\tilde{\phi}_{3}\!)}\label{apprate12proof}
\\&{R_{13}\leq I\!(\!\tilde{d}_{31};\!\tilde{d}_{\!13}\!)\!-\!\!I\!(\!\tilde{d}_{31};\tilde{d}_{\!21},\tilde{d}_{\!23},\tilde{\phi}_{2}\!)}\label{apprate13proof}
\\&{R_{23}\leq I\!(\!\tilde{d}_{23};\!\tilde{d}_{\!32}\!)\!-\!\!I\!(\!\tilde{d}_{23};\tilde{d}_{\!12},\tilde{d}_{\!13},\tilde{\phi}_{1}\!)}\label{apprate23proof}
\end{align}

Each potential eavesdropper combines its available observations to estimate the distance between the other two users to enlarge the subtracted mutual information terms in \eqref{apprate12proof}-\eqref{apprate23proof}. Thus, user $m$ as a potential eavesdropper of the key between users $i$ and $j$ makes estimate of $d_{ij}$ as:
\begin{align}
&\hat{d}_{ij}=\sqrt{\tilde{d}_{mi}^{2}+\tilde{d}_{mj}^{2}-2\tilde{d}_{mi}\tilde{d}_{mj}\cos(\tilde{\phi}_{m})}\label{estimatedij}
 \end{align}
By substituting the parameters inside the square root as \eqref{dij} and \eqref{phii}, we obtain:
\begin{align}
&\hat{d}_{ij}=\sqrt{d_{ij}^{2}+A}\label{estimatedij2}
 \end{align}
in which $A$ is defined in \eqref{estimatedij3} at the top of the next page.
\begin{figure*}[!t]
\small{
\begin{equation}\label{estimatedij3}
A=\!\!N_{\!mi}^{2}\!\!+\!\!N_{\!mj}^{2}\!\!+\!\!2(\!d_{mi}\!N_{\!mi}\!\!+\!\!d_{mj}\!N_{\!mj}\!-\!d_{mi}\!N_{\!mj}\!\cos(\!\tilde{\phi}_{m}\!)\!-\!d_{mj}\!N_{\!mi}\!\cos(\!\tilde{\phi}_{m}\!)\!-\!N_{\!mi}\!N_{\!mj}\!\cos(\!\tilde{\phi}_{m}\!)\!\!+\!\!2d_{mi}d_{mj}\!\sin(\!\frac{N_{\!m}}{2}\!\!)\sin(\!\phi_m\!\!+\!\!\frac{N_{\!m}}{2}\!)\!)
 \end{equation}}
\hrulefill
\end{figure*}

For $J\gg1$, $\sigma_{ij}^{2}/J\ll d_{ij}^{2}$ and $\sigma_{i}^{2}/J\approx 0$, $\forall i\neq j \in \{1,2,3\}$ with high probability. Then \eqref{estimatedij2} can be linearly approximated as:
\begin{align}
&\hat{d}_{ij}\approx d_{ij}(1+\frac{A}{2d_{ij}^{2}})\label{estimatedij4}
 \end{align}
Again by assuming $J\gg1$, $\sigma_{ij}^{2}/J\ll d_{ij}^{2}$ and $\sigma_{i}^{2}/J\approx 0$ and ignoring terms $N_{\!mi}^{2}$, $N_{\!mj}^{2}$ and $N_{\!mi}\!N_{\!mj}\!\cos(\!\tilde{\phi}_{m}\!)$ in $A$ and assuming $\sin(\!\frac{N_{\!m}}{2}\!\!)\approx \frac{N_{\!m}}{2}$, we have
\small{
\begin{align}
&{\hat{d}_{ij}\approx d_{ij}\!+\!\frac{N_{mi}(\!d_{mi}\!-\!d_{mj}\cos(\!\phi_{m}\!)\!)}{d_{ij}}\!+\!\frac{N_{mj}(\!d_{mj}\!-\!d_{mi}\cos(\phi_{m})\!)}{d_{ij}}+}\no
\\&{\ \ \ \ \ \ \ \ \ \ \ \ \ \ \ \ \frac{N_{m}d_{mi}d_{mj}\sin(\!\phi_{m}\!)}{d_{ij}}}\label{estimatedij5}
 \end{align}}
\normalsize{
Since the noise terms in \eqref{estimatedij5} are three independently Gaussian noises, we deduce:}
\begin{align}
{\hat{d}_{ij}\approx d_{ij}+\mathcal{N}(0,\frac{\hat{\sigma}_{ij}^{2}}{J})}\label{hatdij}
 \end{align}
in which:
\begin{align}
&{\hat{\sigma}_{ij}^{2}= \frac{\sigma_{mi}^{2}(\!d_{mi}\!-\!d_{mj}\cos(\!\phi_{m}\!)\!)^{2}}{d_{ij}^{2}}+\frac{\sigma_{mj}^{2}(\!d_{mj}\!-\!d_{mi}\cos(\!\phi_{m}\!)\!)^{2}}{d_{ij}^{2}}+}\no
\\&{\ \ \ \ \ \ \ \ \ \ \ \ \ \ \ \ \frac{\sigma_{m}^{2}(d_{mi}d_{mj}\sin(\!\phi_{m}\!))^{2}}{d_{ij}^{2}}}\label{sigmahatapprox}
 \end{align}
By substituting:
\begin{align}
{\cos(\phi_{m})=\frac{d_{mi}^{2}+d_{mj}^{2}-d_{ij}^{2}}{2d_{mi}d_{mj}}}\no
 \end{align}
and
\begin{align}
{\sin^{2}(\phi_{m})=1-\cos^{2}(\phi_{m})}\no
 \end{align}
in \eqref{sigmahatapprox}, it is rewritten as:
\begin{align}
\hat{\sigma}_{ij}^{2}\!=\!\sigma_{im}^{2}\!+\!\sigma_{jm}^{2}\!+\! \text{Const}_{d_{12},d_{13},d_{23}}(\!\frac{\sigma_{m}^{2}}{4d_{ij}^{2}}\!-\!\frac{\sigma_{im}^{2}}{4d_{ij}^{2}d_{im}^{2}}\!-\!\frac{\sigma_{jm}^{2}}{4d_{ij}^{2}d_{jm}^{2}}\!)
\end{align}
in which
\small{
\begin{align}
&{\text{Const}_{d_{12},d_{13},d_{23}}=(d_{12}+d_{13}+d_{23})\times}\no
\\&{(d_{12}+d_{13}-d_{23})(d_{13}+d_{23}-d_{12})(d_{12}+d_{23}-d_{13})}\no
\end{align}}
\normalsize{
Now, we calculate the bound on rate $R_{12}$, using \eqref{apprate12proof}. The other rates bounds are similarly calculated. We have:}
\begin{align}
&{R_{12}\leq H\!(\!\tilde{d}_{12}|\!\hat{d}_{\!12}\!)\!-\!\!H\!(\!\tilde{d}_{12}|\!\tilde{d}_{\!21}\!)}\label{condent}
\end{align}
Using \eqref{hatdij} and \eqref{dij}, and the fact that at each time slot, $d_{ij}$ is constant, the conditionally Gaussian entropies in \eqref{condent} are calculated. Then, the expected values are computed according to the distributions of $d_{ij}$, $d_{mi}$ and $d_{mj}$.

\subsection*{Proof of Theorem 3}
\noindent To satisfy the rate limitations of the public channels, each user considers a noisy version of its observation to share keys with the other users. We set:
\begin{align}
S_{ij}=\tilde{d}_{ij}+D_{ij}\label{noisyversion1}
\end{align}
\noindent in Theorem 1 where $D_{ij}\!\sim\!{\rm{\mathcal N}}(0,\sigma'^{2}_{ij})$. The noises $D_{ij}$ are independent of each other and of all the observations. In contrast to the unlimited public channel case, the communication between each pair is two-way in general and the total key shared between each pair consists of two keys. By substituting all the auxiliary random variables of Theorem 1 similarly to \eqref{noisyversion1}, it is seen that the sum rates in \eqref{rateregion} and also in the constraints \eqref{cons12}-\eqref{cons123} are inactive. Then the rate region is reduced to:
\small{
\begin{align}
&{R_{12}\leq [I(\tilde{d}_{12}\!+\!D_{12};\!\tilde{d}_{21}\!)\!-\!\!I\!(\!\tilde{d}_{12}\!+\!D_{12};\hat{d}_{12})]^{+}}\no
\\&{+[I(\tilde{d}_{21}\!+\!D_{21};\!\tilde{d}_{\!12}|\tilde{d}_{12}\!+\!D_{12})\!-\!I(\tilde{d}_{21}\!+\!D_{21};\hat{d}_{12}\!|\tilde{d}_{12}\!+\!D_{12})]^{+},}\no
\\&{R_{13}\leq [I(\tilde{d}_{13}\!+\!D_{13};\!\tilde{d}_{31}\!)\!-\!\!I\!(\!\tilde{d}_{13}\!+\!D_{13};\hat{d}_{13})]^{+}}\no
\\&{+[I(\tilde{d}_{31}\!+\!D_{31};\!\tilde{d}_{\!13}|\tilde{d}_{13}\!+\!D_{13})\!-\!I(\tilde{d}_{31}\!+\!D_{31};\hat{d}_{13}\!|\tilde{d}_{13}\!+\!D_{13})]^{+},}\no
\\&{R_{23}\leq [I(\tilde{d}_{23}\!+\!D_{23};\!\tilde{d}_{32}\!)\!-\!\!I\!(\!\tilde{d}_{23}\!+\!D_{23};\hat{d}_{23})]^{+}}\no
\\&{+[I(\tilde{d}_{32}\!+\!D_{32};\!\tilde{d}_{\!23}|\tilde{d}_{23}\!+\!D_{23})\!-\!I(\tilde{d}_{32}\!+\!D_{32};\hat{d}_{23}\!|\tilde{d}_{23}\!+\!D_{23})]^{+},}\no
\end{align}}

\normalsize{
and subject to the constraints:}
%\end{align}
\small{
\begin{align}
&{I\!(\tilde{d}_{12}\!+\!D_{12};\!\tilde{d}_{12}|\tilde{d}_{21})\!+\!I\!(\tilde{d}_{13}\!+\!D_{13};\!\tilde{d}_{13}|\tilde{d}_{31})\!\leq\! R_{1},}\no
\\&{I\!(\tilde{d}_{21}\!+\!D_{21};\!\tilde{d}_{21}|\tilde{d}_{12})\!+\!I\!(\tilde{d}_{23}\!+\!D_{23};\!\tilde{d}_{23}|\tilde{d}_{32})\!\leq\! R_{2},}\no
\\&{I\!(\tilde{d}_{31}\!+\!D_{31};\!\tilde{d}_{31}|\tilde{d}_{13})\!+\!I\!(\tilde{d}_{32}\!+\!D_{32};\!\tilde{d}_{32}|\tilde{d}_{23})\!\leq\! R_{3}.}\no
\end{align}}

\normalsize{
By the same arguments as in the unlimited public channel case and the same calculations, the rate region in Theorem 3 is deduced.

\section*{Appendix C}

\noindent
\subsection*{Proof of Corollary 1}
\noindent We use the following explicit outer bound on the pairwise key capacity region, given in \cite{salimi-pairwise}, which is based on unlimited public channel:
\begin{align}
&{R_{12}\leq I(X_{1};X_{2}|X_{3}),}\no
\\&{R_{13}\leq I(X_{1};X_{3}|X_{2}),}\no
\\&{R_{23}\leq I(X_{2};X_{3}|X_{1}).}\no
\end{align}

We calculate the upper bound on $R_{12}$ and similarly, the other upper bounds can be concluded. We have:

$\begin{array}{l}{I(X_{1};X_{2}|X_{3})\mathop{=}\limits^{(a)}I(\tilde{d}_{12};\tilde{d}_{21}|\tilde{d}_{31},\tilde{d}_{32},\tilde{\phi}_{3})}
\\ {=H(\tilde{d}_{12}|\tilde{d}_{31},\tilde{d}_{32},\tilde{\phi}_{3})-H(\tilde{d}_{12}|\tilde{d}_{21},\tilde{d}_{31},\tilde{d}_{32},\tilde{\phi}_{3})}
\\ {\leq H(\tilde{d}_{12}|\tilde{d}_{31},\tilde{d}_{32},\tilde{\phi}_{3})-H(\tilde{d}_{12}|d_{12},\tilde{d}_{21},\tilde{d}_{31},\tilde{d}_{32},\tilde{\phi}_{3})}
\\ {\mathop{=}\limits^{(b)} H(\tilde{d}_{12}|\tilde{d}_{31},\tilde{d}_{32},\tilde{\phi}_{3})-H(\tilde{d}_{12}|d_{12})}
\\ {= H(\tilde{d}_{12}|\tilde{d}_{31},\tilde{d}_{32},\tilde{\phi}_{3})-\frac{1}{2}\log(2\pi e\sigma_{12}^{2})}
\\ {\mathop{=}\limits^{(c)} H(\tilde{d}_{12}|\hat{d}_{12},\tilde{d}_{31},\tilde{d}_{32},\tilde{\phi}_{3})-\frac{1}{2}\log(2\pi e\sigma_{12}^{2})}
\\ {\leq H(\tilde{d}_{12}|\hat{d}_{12})-\frac{1}{2}\log(2\pi e\sigma_{12}^{2})}
\\ {= H(\tilde{d}_{12}-\hat{d}_{12}|\hat{d}_{12})-\frac{1}{2}\log(2\pi e\sigma_{12}^{2})}
\\ {\leq H(\tilde{d}_{12}-\hat{d}_{12})-\frac{1}{2}\log(2\pi e\sigma_{12}^{2})}
\\ {\mathop{\leq}\limits^{(d)} \frac{1}{2}\log(2\pi e(\sigma_{12}^{2}+\mathbb{E}(\hat{\sigma}_{12}^{2})))-\frac{1}{2}\log(2\pi e\sigma_{12}^{2})}
\vspace{.2cm}
\\ {=\frac{1}{2}\log(1+\frac{\mathbb{E}(\hat{\sigma}_{12}^{2})}{\sigma_{12}^{2}})}
\end{array}$

\noindent In the above equations, (a) follows from the fact that distances $d_{12}$, $d_{13}$  and $d_{23}$ and also the respective noises are independent of each other. (b) is true since $\tilde{d}_{12}-d_{12}-(\tilde{d}_{21},\tilde{d}_{31},\tilde{d}_{32},\tilde{\phi}_{3})$. (c) is due to the fact that $\hat{d}_{12}$ is a function of  $(\tilde{d}_{31},\tilde{d}_{32},\tilde{\phi}_{3})$. (d) is deduced with the argument that for a given variance, Gaussian distribution maximizes the entropy. To calculate the entropy of $\tilde{d}_{12}-\hat{d}_{12}$, we use the following formula:
\begin{align}
&{\mathbb{V}ar(\tilde{d}_{12}-\hat{d}_{12})=\mathbb{E}(\mathbb{V}ar(\tilde{d}_{12}-\hat{d}_{12}|d_{12},d_{13},d_{23}))+}\no
\\&{\mathbb{V}ar(\mathbb{E}(\tilde{d}_{12}-\hat{d}_{12}|d_{12},d_{13},d_{23})).}\no
\end{align}
Since
\begin{equation}\no
\mathbb{E}(\tilde{d}_{12}-\hat{d}_{12}|d_{12},d_{13},d_{23})=0,
\end{equation}
and
\begin{equation}\no
\mathbb{V}ar(\tilde{d}_{12}-\hat{d}_{12}|d_{12},d_{13},d_{23})=\sigma_{12}^{2}+\hat{\sigma}_{12}^{2},
\end{equation}
 we have:
\begin{equation}
\mathbb{V}ar(\tilde{d}_{12}-\hat{d}_{12})=\sigma_{12}^{2}+\mathbb{E}(\hat{\sigma}_{12}^{2})
\end{equation}
and then, the outer bound in Corollary 1 is deduced.

}

%\begin{thebibliography}{1}
%
%
%\bibitem{CoverBok} T. M. Cover and J. A. Thomas,  \emph{Elements of Information Theory}, Wiley Series in Telecommunications and Signal Processing, Wiley-Interscience 2006.
%
%
%\bibitem{cover_relay} T. M. Cover and A. El Gamal, \lq\lq Capacity theorems for the relay
%channel,\rq\rq ~\emph{IEEE Trans. Inf. Theory}, vol. 25, no. 5,
%pp. 572-584, Sep. 1979.
%
%\end{thebibliography}

}
%
%
%In subsections there is 1 blank line before the section
%heading and one afterwards.  Heading text is 11 point
%bold font.  Paragraphs are indented one pica.  There is
%no blank line between paragraphs.
%Throughout I may cite references of the form
%\cite{key:foo} or \cite{foo:baz}, and LaTeX will keep
%track of numbering.  The numbers are based on the order
%you place them in the bibliography, not the order they
%appear in the text.  They should (I believe) be in
%alphabetical order.  LaTex will put square brackets about
%the number within the text of your paper.  For those of
%you new to the bibliography package, you may have to run
%the latex process twice to allow all references to be
%resolved. You will get a warning about a missing .aux
%file.  Just rerun latex and it will be ok.
%\section{Summary and Conclusions}
%This template will get you through the minimum article,
%i.e. no figures or equations.  To include those, please
%refer to your LaTeX manual and the IEEE publications
%guidelines.  Good Luck!
%%this is how to do an unnumbered subsection
%\subsection*{Acknowledgments}
%This is how to do an unnumbered subsection, which comes
%out in 11 point bold font.  Here I thank my colleagues,
%especially Mike Gennert, who know more about Tex and
%Latex than I.
%\begin{thebibliography}{9}
%\bibitem{key:foo}
%I. M. Author,
%``Some Related Article I Wrote,''
%{\em Some Fine Journal}, Vol. 17, pp. 1-100, 1987.
%\bibitem{foo:baz}
%A. N. Expert,
%{\em A Book He Wrote,}
%His Publisher, 1989.
%\end{thebibliography}
\end{document}